\documentclass[12pft]{article}
\usepackage{latexsym,amsmath,amssymb,}
\usepackage{graphicx}
\usepackage{ulem}
\usepackage{dsfont}

\begin{document}
\title{Time Evolution of the Radial Perturbations  and Linear Stability of Solitons and Black Holes in a Generalized Skyrme Model}

\author{Daniela D. Doneva$^{1,2}$\thanks{E-mail: ddoneva@phys.uni-sofia.bg}, Kostas D. Kokkotas$^{2}$ \thanks{E-mail: kostas.kokkotas@uni-tuebingen.de} \,\,
\\{\footnotesize  ${}^{1}$Deptartment of Astronomy,
                Faculty of Physics, St.Kliment Ohridski University of Sofia}\\
                {\footnotesize  5, James Bourchier Blvd., 1164 Sofia, Bulgaria }\\\\[-3.mm]
      {  \footnotesize ${}^{2}$ Theoretical Astrophysics, Eberhard-Karls University of T\"ubingen, T\"ubingen 72076, Germany }\\\\[-3.mm]
  Ivan Zh. Stefanov$^{3}$ \thanks{E-mail: izhivkov@tu-sofia.bg}, Stoytcho S. Yazadjiev$^{4}$ \thanks{E-mail: yazad@phys.uni-sofia.bg}
 \\ [2.mm]{\footnotesize{${}^{3}$ Department of Applied Physics, Technical University of Sofia,}}\\ [-1.mm]{\footnotesize{8, Kliment Ohridski Blvd.,
 1000
Sofia, Bulgaria}}\\\\[-3.mm]
      {  \footnotesize ${}^{4}$Department of Theoretical Physics,
                Faculty of Physics, St.Kliment Ohridski University of Sofia}\\
{\footnotesize  5, James Bourchier Blvd., 1164 Sofia, Bulgaria }\\[-3.mm]}

\date{}

\maketitle

\begin{abstract}
We study the time evolution of the radial perturbation for self-gravitating soliton and black-hole solutions in a
generalized Skyrme model in which a dilaton is present. The background solutions were obtained recently by some of the authors.
For both the solitons and the black holes two branches of solutions exist which merge at some critical
value of the corresponding parameter. The results show that, similar to the case without a scalar field,
one of the branches is stable against radial perturbations and the other is unstable. The conclusions for the linear
stability of the black holes in the generalized Skyrme model are also in agreement with the results from the
thermodynamical stability analysis based on the turning point method.
\end{abstract}


\sloppy
\section{Introduction}
Our intuition for the properties of the solutions describing self-gravitating objects in general relativity is based, to a large extent, on some exact
solutions which belong to the Kerr-Newman class of black holes. For these solutions uniqueness theorems,
and theorems stating that globally regular self-gravitating solutions (solitons) do not exist, have been proven rigorously in the case of
vacuum or linear matter models such as Maxwell electrodynamics \cite{Carter}--\cite{Heusler}. As the investigations in the last two decades revealed, the standard intuition
often fails when nonlinear matter models are considered \cite{Bizon94}, which makes the study of self-gravitating solutions in such models vital
for fundamental physics.

One of the effective nonlinear matter models which has attracted much attention is Skyrme's theory \cite{Skyrme61},\cite{SkyrmeBook}.
In this theory baryons are described as solitons
in an effective theory of mesons. The interest in Skyrme theory was revived in the 1980s when it was found that the Skyrme Lagrangian can be
derived from quantum chromodynamics (QCD) in the low-energy regime.

Self-gravitating solutions in Skyrme theory were considered for the first time by Luckock et al. \cite{Luckock}. The solutions
in Einstein-Skyrme (ES) theory are nonunique, and those with a nontrivial Skyrme field can be divided into two branches. The first branch of
solutions has a well-defined flat-space limit. It was obtained by
Droz, Heusler  and Straumann \cite{DrozHeusler91sol}. The authors found that these solutions are stable against spherically symmetric
perturbations \cite{HeuslerDroz91stab, HeuslerDrozStabLin}. The second branch of solutions was discovered by Bizon and Chmaj soon after
that \cite{Bizon92}. This branch has no flat-space limit and it is unstable, as the authors' analysis revealed.
The stability of the ES solitons has also been studied in \cite{Bizon07, Zajac09, Zajac10}.
There is also a branch of solutions that has a trivial Skyrme field and coincides with the pure Schwarzschild black hole.
Self-gravitating solutions in Skyrme theory, both black holes and solitons, have also
been studied in a series of papers \cite{Glendenning88}--\cite{Piette07}.

Different modifications of Skyrme theory have been considered in order to cure some of its deficiencies which are present in the original
version of the theory \cite{Skyrme61}. One possible generalization is the
inclusion of a dilaton. The dilaton is added in the theory to restore scale invariance which is also characteristic for the underlying QCD.
It has also been considered as a source of additional intermediate-range attractive forces which are vital for the formation of stable multisoliton
configurations such as nuclei and baryon stars. A Generalized Skyrme Model (GSM) which includes a dilaton has been derived from QCD in the
low-energy regime in \cite{Adrianov86, Adrianov87, Nikolaev92}. In a recent paper \cite{DSY_SKYRME} we reported numerical solutions describing
self-gravitating solitons and black holes in the GSM. They are generalizations of the soliton and black-hole solutions that have been obtained numerically
in \cite{DrozHeusler91sol} and \cite{Bizon92}.

The aim of the current paper is to study the response of the self-gravitating GSM solutions \cite{DSY_SKYRME}, both soliton and black-hole types, to
small radial perturbations and, in particular, to determine if the inclusion of the dilaton in Skyrme theory changes the stability properties.
We study the quasinormal modes (QNMs) of the solutions by evolving the time-dependent wave equations.

The problem of studying the QNMs and the stability of the GSM solutions is mathematically more complex than that of the ES solutions since in the
former case a system of two coupled wave equations for the perturbations of the Skyrme field and the dilaton has to be
solved even though the considerations are restricted to radial perturbations (in the ES case the problem is reduced to only one wave equation
for the Skyrme field).
What makes the problem even more difficult is that the wave equation for the perturbations of the dilaton contains a potential
which is not vanishing at infinity, i.e. the scalar field is massive and the time evolution of the perturbations
has some specific properties \cite{Hod}--\cite{Konoplya2005}.

The paper is organized as follows. The GSM coupled to gravity is briefly presented in Section \ref{sec:model}. In this section the time-dependent field
equations are given. The system of coupled equations for the radial part of the perturbations of the Skyrme field and the dilaton is derived in
Section \ref{sec:perturb} and solved with the proper boundary conditions numerically in Section
\ref{sec:Results}. In Section \ref{sec:Conclusion} a summary of the results is given.

\section{The Generalized Skyrme Model}\label{sec:model}
\subsection*{Action}
Let us briefly introduce the model considered in \cite{DSY_SKYRME}. We start with the following action:
\begin{equation}
S=\int{d^4x \sqrt{-g} \left(-\frac{R}{16\pi G} + L_M \right)}. \label{eq:action}
\end{equation}
The flat-space Lagrangian of the GSM can be found in \cite{Nikolaev92}. When gravity is included the GSM Lagrangian is naturally generalized to the form
\begin{eqnarray}
&&L_M = \frac{1}{4} f_\pi^2~ \exp(-2\sigma)~ {\rm Tr}[\nabla_\mu U \nabla^\mu U^+]  + \frac{N_f f_\pi^2}{4}~ \exp(-2\sigma)~ g^{\mu\nu} \partial_\mu \sigma
\partial_\nu \sigma
\label{eq:Lagrang}\\ \notag \\
&& \hspace{0.5cm} + \frac{1}{32 e^2} {\rm Tr}[(\nabla_\mu U) U^{+},(\nabla_\nu U)
U^{+}]^2  + V_{\rm GSM}(\sigma), \notag
\end{eqnarray}
where the derivatives have been substituted with covariant derivatives.
Here $U$ is the SU(2) chiral field, $\sigma$ is the dilaton, $\nabla_\mu$ is the covariant derivative with respect to the metric
$g_{\mu\nu}$, $f_\pi$ is the pion decay
constant, $e$ is the Skyrme constant, $C_g$ is the gluon condensate, $N_f$ is the number of flavors, and $\varepsilon =
8N_f/(33-2N_f)$.
The first two terms in (\ref{eq:Lagrang}) are the kinetic terms for the chiral and the dilaton fields. The third term is the
one introduced by Skyrme for the stabilization of the soliton solutions. The potential of the dilaton field is given by
\begin{equation}
V_{\rm GSM}(\sigma) = -\frac{C_g N_f}{48} \left[ \exp(-4\sigma) - 1 + \frac{4}{\varepsilon} (1-\exp(-\varepsilon
\sigma))
\right]. \label{eq:PotentOrig}
\end{equation}
The dilaton couples only to those terms of  Lagrangian density that break the scale invariance\footnote{For more details we refer
the reader to \cite{DSY_SKYRME}.}.

Instead of $\sigma$ it is more convenient to work with the function $\Phi$ which is defined by
\begin{equation}
\Phi=\exp(-\sigma).
\end{equation}

\subsection*{Reduced Lagrangian}

We are going to restrict our considerations to the spherically symmetric case.
In \cite{DSY_SKYRME} the hedgehog ansatz for the chiral field
\begin{equation}
U=\exp[\mathbf{\tau}\cdot \hat{\mathbf{r}} F(r,t)]
\end{equation}
was chosen. Here $\tau$ are the Pauli matrices and $\hat{\mathbf{r}}$ is a unit radial vector.
With the following time-dependent ansatz for the metric,
\begin{equation}
ds^2 = e^{\chi(t,r)} dt^2 - e^{\alpha(t,r)} dr^2 - r^2(d\theta^2 + \sin^2{\theta}
d\varphi^2) \label{eq:metric}
\end{equation}
the Lagrangian (\ref{eq:Lagrang}) takes the form\footnote{The notation we choose here is slightly different from that in \cite{DSY_SKYRME}.
It facilitates the comparison of the equations and the results to the ES case \cite{DrozHeusler91sol}--\cite{Bizon92}.  }
\begin{equation}
L_m={a^2\over b} \left[{u\over x^2}\left(e^{-\chi}\dot{F}^2-e^{-\alpha}F\,'\,^2\right)-{v\over x^2}
+\tilde{N}\left(e^{-\chi}\dot{\Phi}^2-e^{-\alpha}\Phi\,'\,^2\right)+{1\over a} \tilde{V}\right],\label{eq:Lagr_red}
\end{equation}
where
\begin{eqnarray}
&&u=x^2 \Phi^2+2 \sin^2F, ~~v=\left(2\Phi^2+{\sin^2F\over x^2}\right)\sin^2F, \\
&& \tilde{V}(\Phi)=\frac{16 \pi G b}{a}V_{\rm GSM}(\Phi)=-\frac{\gamma \tilde{N} b}{a}\left[\Phi^4-1+{4\over
\varepsilon}\left(1-\Phi^{\varepsilon}\right)\right]. \label{eq:PotentPhi}
\end{eqnarray}
We have introduced the following constants:
\begin{equation}
a = 8\pi G f^2_{\pi}, ~~ b=8\pi G \frac{1}{e^2}, ~~ \gamma=2 \pi G \frac{C_g}{3}, ~~\tilde{N}={N_f\over2},
\end{equation}
and dimensionless variables  $\tau = e f_\pi t$, $x = e f_\pi r$.
The derivative with respect to the dimensionless time coordinate $\tau$ is denoted by a dot, while the derivative with respect to the
dimensionless radial coordinate $x$ is denoted by a prime.
Below we will also use the parameter\footnote{The parameter $a$ is two times bigger that the parameter $\alpha$ used
in \cite{DrozHeusler91sol}--\cite{Bizon92} and the parameter $D_{\rm eff}$ is chosen to be the same as in \cite{Nikolaev92}.}

\begin{equation}
D_{\rm eff} = { \gamma \tilde{N}\over 2~a~ e^2 f_\pi^2 }. \label{eq:Deff}
\end{equation}
For the number of flavors, we fixed the value  $N_f=2$, so $\tilde{N}=1$.
\subsection*{Time-dependent field equations}
The Einstein equations have the following form:
\begin{equation}
G_{\mu\nu}=-{1\over 2}T_{\mu\nu},\label{eq:Einst}
\end{equation}
\begin{equation}
T_{\mu\nu}=-g_{\mu\nu}L_m+2{\delta L_m\over \delta g^{\mu\nu}}.
\end{equation}
The $(tt)$, $(rr)$, and $(tr)$ components of (\ref{eq:Einst}) are
\begin{equation}
\left[e^{-\alpha}\left(1-x\alpha'\right)-1\right]=-a\left({1\over 2}u w+{1\over 2} v+{1\over 2}x^2 z\right)+{1\over 2}x^2\tilde{V},\label{eq:Einst_tt}
\end{equation}
\begin{equation}
\left[e^{-\alpha}\left(1+x\chi'\right)-1\right]=a\left({1\over 2}u w-{1\over 2} v+{1\over 2}x^2 z\right)+{1\over 2}x^2\tilde{V},\label{eq:Einst_rr}
\end{equation}
\begin{equation}
\dot{\alpha}={a\over x}\left(u \dot{F}F\,'+x^2 \tilde{N}\dot{\Phi}\Phi\,'\right),\label{eq:Einst_tr}
\end{equation}
where
\begin{eqnarray}
w&=&e^{-\chi}\dot{F}^2+e^{-\alpha}F\,'\,^2,\\
z&=&\tilde{N}\left(e^{-\chi}\dot{\Phi}^2+e^{-\alpha}\Phi\,'\,^2\right).
\end{eqnarray}
The combination of equations (\ref{eq:Einst_tt}) and (\ref{eq:Einst_rr}) gives the following useful expression:
\begin{equation}
{\chi\,'-\alpha\,'\over2}={e^{\alpha}\over x} \left(1-{1\over2}a\, v+{1\over2}x^2 \tilde{V}\right)-{1\over x}\,\,.\label{eq:Einst_tt_rr}
\end{equation}
The time-dependent field equations for $F$ and $\Phi$ obtained from (\ref{eq:Lagr_red}) are
\begin{multline}\label{eq:F}
e^{\alpha-\chi}\left[{\dot{\alpha}-\dot{\chi}\over2}u \dot{F}+(u \dot{F})\dot{}\right]=\left[{\chi'-\alpha'\over2}u F\,'+(u F\,')\,'\right]\\
+{1\over 2}u_F\left(e^{\alpha-\chi}\dot{F}^2-F\,'\,^2\right)-{1\over 2}e^{\alpha}v_F=0,
\end{multline}
\begin{multline}\label{eq:Phi}
e^{\alpha-\chi}\left[{\dot{\alpha}-\dot{\chi}\over2}x^2 \dot{\Phi}+(x^2 \dot{\Phi})\dot{}\right]=\left[{\chi'-\alpha'\over2}x^2 \Phi'+(x^2 \Phi')\,'\right]\\
+{1\over 2\tilde{N}}u_{\Phi}\left(e^{\alpha-\chi}\dot{F}^2-F\,'\,^2\right)-{1\over 2\tilde{N}}e^{\alpha}v_{\Phi}+{x^2\over
2a\tilde{N}}e^{\alpha}\tilde{V}_{\Phi}=0,
\end{multline}
where  $(..)_\Phi$  denotes the partial derivative with respect to $\Phi$, and  $(..)_F$ denotes the partial derivative with respect to $F$.

\section{Equations for the radial perturbations}\label{sec:perturb}
We reduce our considerations to radial perturbations
\begin{eqnarray}
\alpha(\tau,x)=\alpha_0(x)+\delta \alpha(\tau,x),\notag\\
\chi(\tau,x)=\chi_0(x)+\delta \chi(\tau,x),\notag\\
F(\tau,x)=F_0(x)+\delta F(\tau,x),\notag\\
\Phi(\tau,x)=\Phi_0(x)+\delta \Phi(\tau,x),\notag
\end{eqnarray}
and follow the scheme presented in \cite{HeuslerDroz91stab}.
It turns out that the evolution of the Skyrme field and the scalar field perturbations, $\delta F$ and $\delta \Phi$, respectively, can be
studied independently from the perturbations of the metric.
The equation for $\delta F$, obtained from (\ref{eq:F}), is
\begin{multline}
e^{\alpha_0-\chi_0}u_0\ddot{\delta F}=u_0\delta F\,'' + \left({\chi_0\,'-\alpha_0\,'\over2}u_0+u_0'\right)\delta F\,'+u_{0\Phi} F_0^{\,'}\delta \Phi\,'+\\u_{0\Phi}' F_0^{\,'}\delta \Phi+
\left(F_0^{\,''}+{\chi_0\,'-\alpha_0\,'\over2}F_0^{\,'}\right)\delta u+u_{0F}'F_0^{\,'}\delta F-{1\over2}F_0^{\,'2}\delta u_F-\\{1\over2}e^{\alpha_0}\delta v_F-{1\over2}e^{\alpha_0}v_{0F}\delta\alpha+u_0F_0^{\,'}{\delta\chi\,'-\delta\alpha\,'\over2}. \label{eq:var_F}
\end{multline}
From (\ref{eq:Phi}) we obtain the following equation for the perturbations of the scalar field $\delta \Phi$
\begin{multline}
e^{\alpha_0-\chi_0}x^2\ddot{\delta \Phi}=x^2\delta \Phi\,'' + \left({\chi_0\,'-\alpha_0\,'\over2}x^2+2x\right)\delta \Phi\,'-{1\over \tilde{N}}u_{0\Phi} F_0^{\,'}\delta F\,'\\-
{1\over 2\tilde{N}}F_0^{\,'2} \delta u_{\Phi}+ {1\over 2\tilde{N}} e^{\alpha_0} \left(-\delta v_{\Phi}+{x^2\over a}\delta \tilde{V}_{\Phi}\right) +
\\{1\over 2\tilde{N}} e^{\alpha_0} \left(-v_{0\Phi}+{x^2\over a} \tilde{V}_{0\Phi}\right)\delta\alpha+x^2\Phi_0'{\delta\chi\,'-\delta\alpha\,'\over2},\label{eq:var_Phi}
\end{multline}
where
$$\delta u=u_{0F}\delta F +u_{0\Phi}\delta\Phi, \,\,\,\,\,\,\,\,\,\,\delta v=v_{0F}\delta F +v_{0\Phi}\delta\Phi,$$
$$\delta u_F=u_{0FF}\delta F +u_{0F\Phi}\delta\Phi, \,\,\,\,\,\,\,\,\,\,\delta v_F=v_{0FF}\delta F +v_{0F\Phi}\delta\Phi,$$
$$\delta u_{\Phi}=u_{0\Phi F}\delta F +u_{0\Phi\Phi}\delta\Phi, \,\,\,\,\,\,\,\,\,\,\delta v_{\Phi}=v_{0\Phi F}\delta F +v_{0\Phi\Phi}\delta\Phi,\,\,\,\,\,\,\,\,\,\,\delta\tilde{V}_{\Phi}=\tilde{V}_{0\Phi\Phi}\delta\Phi.$$\\
Throughout the paper, the lower index $(..)_0$ means that the corresponding quantity
refers to the background static solution.
Lower indices $F$ and $\Phi$ denote the corresponding partial derivatives.
The variation of eq. (\ref{eq:Einst_tr}) gives
\begin{equation}
\delta\dot{\alpha}={a\over x}\left(u_0 F_0^{\,'}\dot{\delta F}+x^2 \tilde{N}\Phi_0'\dot{\delta\Phi}\right).
\end{equation}
The integration of the above expression with respect to $\tau$ gives
\begin{equation}
\delta\alpha={a\over x}\left(u_0 F_0^{\,'}\delta F+x^2\tilde{N}\Phi_0'\delta\Phi\right)\label{eq:relation2}
\end{equation}
and it allows us to relate the perturbations of the metric functions $\delta \alpha$
to the perturbations of the matter fields $\delta F$ and $\delta \Phi$.
Another useful relation can be obtained from (\ref{eq:Einst_tt_rr})
\begin{equation}
{\delta\chi\,'-\delta\alpha\,'\over2}={e^{\alpha_0}\over x} \left[\left(1-{1\over2}a\, v_0+{1\over2}x^2 \tilde{V}_0\right)\delta\alpha-{1\over2}a\, \delta v+{1\over2}x^2 \delta\tilde{V}\right].\label{eq:relation3}
\end{equation}
Relations (\ref{eq:relation2})--(\ref{eq:relation3}), substituted back into (\ref{eq:var_F}) and (\ref{eq:var_Phi}), allow us to exclude the variations
of the metric and to obtain a system of two coupled equations (each of them of second order) for $\delta F$ and $\delta \Phi$.

By the following substitution,
\begin{equation}
\delta F={\zeta\over \sqrt{u_0}}, \quad\quad \delta \Phi={\Psi\over x},
\end{equation}
we obtain a system of coupled wave equations
\begin{eqnarray}
-e^{\alpha_0-\chi_0}\ddot{\zeta}+\zeta'' + {\chi_0\,'-\alpha_0\,'\over2}\zeta'+A_1\zeta+A_2\Psi'+A_3\Psi=0,\label{eq:wave_like_zeta}\\
-e^{\alpha_0-\chi_0}\ddot{\Psi}+\Psi'' + {\chi_0\,'-\alpha_0\,'\over2}\Psi'+B_1\Psi+B_2\zeta'+B_3\zeta=0\label{eq:wave_like_psi}.
\end{eqnarray}
The coefficients $A_1$, $A_2$, $A_3$,$B_1$, $B_2$, and $B_3$ are given in  Appendix \ref{sec:coefficients}.
If we multiply (\ref{eq:wave_like_zeta}) and (\ref{eq:wave_like_psi}) by $e^{\chi_0-\alpha_0}$ and introduce a new radial coordinate
\begin{eqnarray}
d x_*={d x\over e^{(\chi_0-\alpha_0)/ 2}},
\end{eqnarray}
the system of equations takes the form
\begin{eqnarray}
-{\partial^2\zeta\over \partial\tau^2}+{\partial^2\zeta\over \partial x_*^2} + \tilde{A}_1\zeta+\tilde{A}_2{\partial\Psi\over \partial x_*}+\tilde{A}_3\Psi=0,\label{eq:wave_zeta}\\
-{\partial^2\Psi\over \partial\tau^2}+{\partial^2\Psi\over \partial x_*^2}+\tilde{B}_1\Psi+\tilde{B}_2{\partial\zeta\over \partial x_*}+\tilde{B}_3\zeta=0\label{eq:wave_psi}.
\end{eqnarray}
The coefficients $\tilde{A}_1$, $\tilde{A}_2$, $\tilde{A}_3$,$\tilde{B}_1$, $\tilde{B}_2$, and $\tilde{B}_3$ are given in  Appendix \ref{sec:coefficients}.

Let us describe the qualitative properties of the wave equations. If we had one wave equation
(which could be transformed to a stationary Schr\"{o}dinger-like equation by a proper separation of the time and spatial variables)  the QNM frequencies would depend strongly on the shape of the potential.
In our case we have two coupled wave equations and the notion of potential is not so clear. Still, the coefficients in front of the zeroth
order derivatives of the wave functions, $\tilde{A}_1$ and $\tilde{B}_1$, respectively, determine the properties of the solutions.
Thus we introduce two functions  $U^{\zeta}$ and $U^{\Psi}$ which can be expressed by $\tilde{A}_1$ and $\tilde{B}_1$ and which we will call potentials of the wave equations for $\zeta$ and $\Psi$, respectively. As the results show, the presence of unstable
modes (solutions divergent with time), depends on whether these potentials have a deep enough negative minimum.

The potentials $U^{\zeta}$ and $U^{\Psi}$ for the solitons are defined as
\begin{equation}
U^{\zeta}=-e^{\chi_0-\alpha_0}\left(A_1-{2\over x^2}\right)=-\left(\tilde{A}_1-{2\,e^{\chi_0-\alpha_0}\over x^2}\right),
\quad\quad U^{\Psi}=-\tilde{B}_1, \label{eq:Potentials_Solitons}
\end{equation}
where $U^{\zeta}$ is chosen in such a way  that it is regular at the origin (see \cite{HeuslerDroz91stab} for a more
detailed discussion on that definition) and $U^{\Psi}$ is also finite at $x=0$.
The potentials for the black-hole solutions are simply
\begin{equation}
U^{\zeta}=-\tilde{A}_1, \quad\quad U^{\Psi}=-\tilde{B}_1, \label{eq:Potentials_BH}
\end{equation}
where both $U^{\zeta}$ and $U^{\Psi}$ are zero on the event horizon $x_H$.

It is also important to comment on the asymptotic value of the potentials at infinity. For both the black holes and the
solitons, $U^{\zeta}$ tends to zero when $x\rightarrow \infty$ , but
$U^{\Psi}$ has a nonzero value at infinity. The reason is that the scalar field we are considering is
massive because of the specific form of the potential $\tilde{V}(\Phi)$, defined by eq. (\ref{eq:PotentOrig}). The mass $m$ of the scalar field
can be defined through the asymptotic value of the potential $U^{\Psi}$, i.e.
as
\begin{equation}
\lim_{x \rightarrow \infty} U^{\Psi} = - m^2.
\end{equation}
Using equation (\ref{eq:var_Phi}) it can be easily derived that
\begin{equation}
\lim_{x \rightarrow \infty} \left(\frac{\tilde{V}_{0\,\Phi\Phi}}{2\tilde{N}a}\right) = -m^2.
\end{equation}

One of the main differences in the time evolution of a massive test scalar field in comparison
to the massless case, is that the tail is oscillating with the period  \cite{Hod}
\begin{equation}
T={2\pi\over m}. \label{eq:PeriodTail}
\end{equation}
In the limit $m\rightarrow 0$ the oscillations of the tail disappear and we are left with the standard power law
tail. Even though in our problem we are dealing with two wave equations -- one for the Skyrme field and one for the massive scalar
field -- it is expected (and confirmed by the numerical results) that the tail will again be oscillatory with period (\ref{eq:PeriodTail}).

Another point worth mentioning is the qualitative behavior of the QNM frequencies for the stable and the unstable
modes. As it is well known, the frequencies of the stable modes are complex, where the real part is inversely proportional to the period of the
oscillations and the imaginary -- to the damping time. The picture changes when the modes are unstable. In this
case the frequencies are purely imaginary, i.e. there is no oscillation and the modes grow exponentially with time
\cite{Doneva2010}, \cite{ThorneBook}.

\section{Numerical results}\label{sec:Results}

\subsection{Solitons}\label{sec:Solitons}

\begin{figure}
    \begin{center}
    \includegraphics[width=7.5cm]{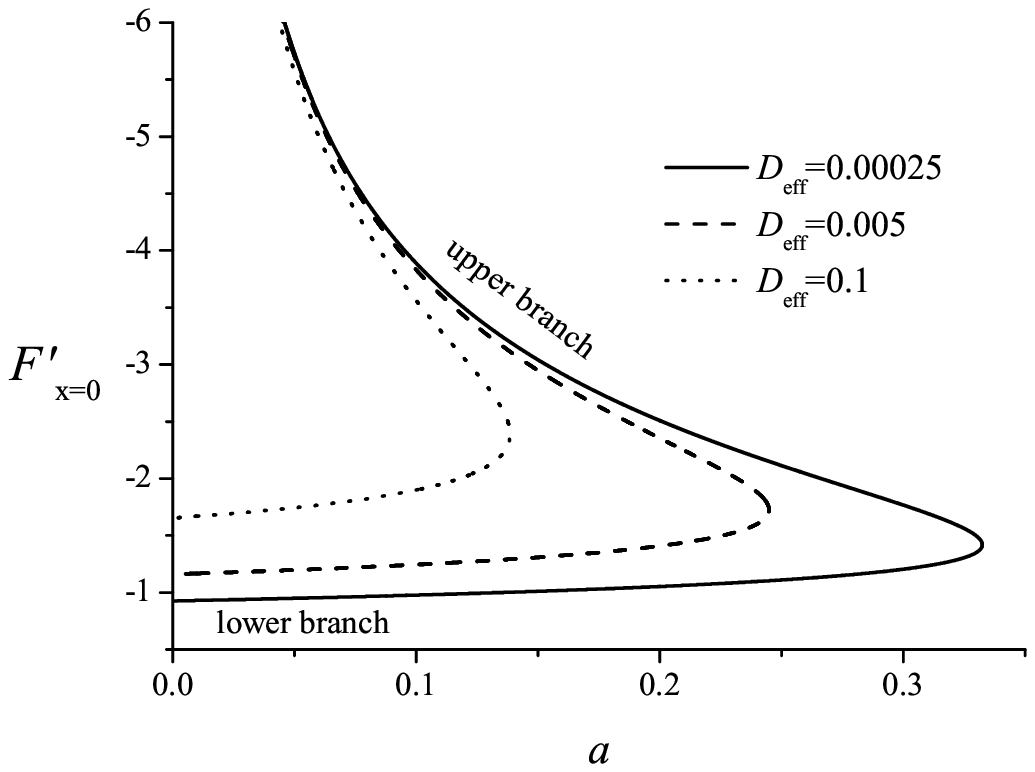}
    \includegraphics[width=7.5cm]{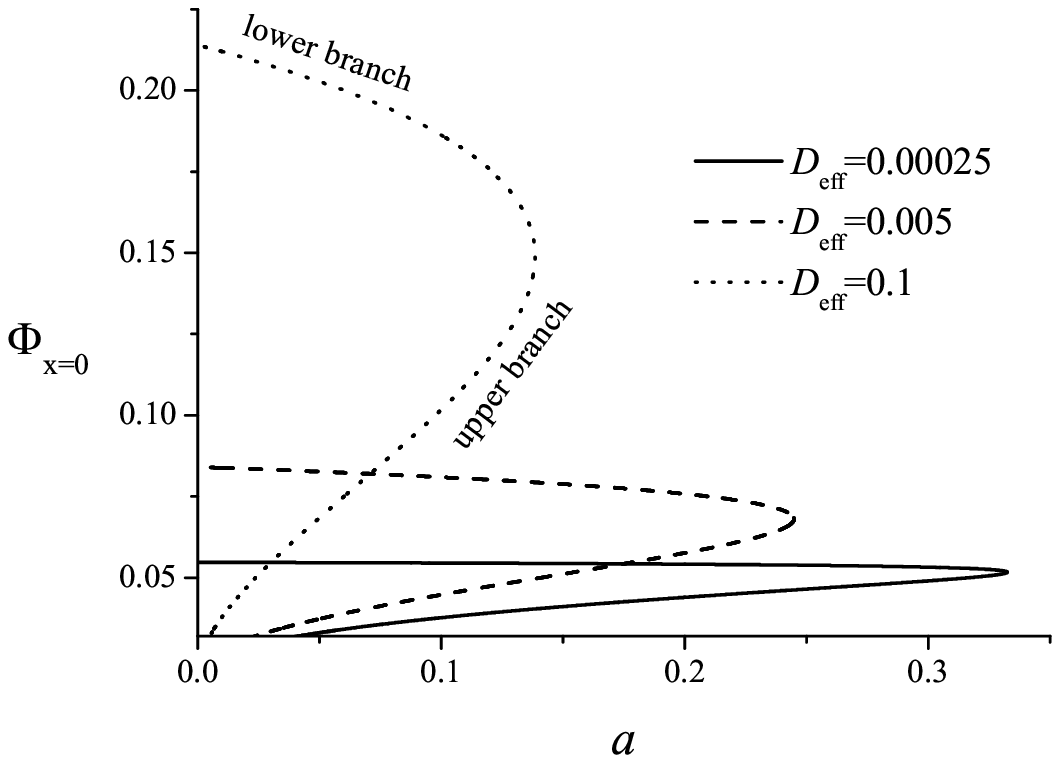}
    \caption{
      The $F_{x=0}^{\,'}(a)$ and $\Phi_{x=0}(a)$ phase diagrams for sequences of soliton solutions for $n=1$ and different values of $D_{\rm eff}$.}
    \label{fig:phase_soliton}
\end{center}
\end{figure}

The background soliton solutions
have been obtained in \cite{DSY_SKYRME}. These solutions are topologically nontrivial and the integer $n$ that occurs in the boundary condition for
the Skyrme field at the origin $F_{x=0}=n \pi$ is interpreted as the baryon number. Once $n$ is fixed, the soliton
solutions obtained in \cite{DSY_SKYRME} are labeled by the values of the shooting parameters $F_{x=0}^{\,'}$ and $\Phi_{x=0}$, where
the index $(..)_{x=0}$ refers to the value of the function calculated at the origin $x=0$. An example of the $F_{x=0}^{\,'}(a)$ and $\Phi_{x=0}(a)$
phase diagrams, presenting sequences of soliton solutions for $n=1$, is shown in Fig. \ref{fig:phase_soliton}.
From the figure it can be seen that the solutions are divided into two branches -- the so-called upper and lower branches --
and the two branches merge at some critical value of the parameter $a_{\rm crit}$
\footnote{There is a discrete infinite series of copies of these branches corresponding to higher excitations with $n>0$ \cite{DSY_SKYRME} but they
will not be discussed here since they are energetically unstable \cite{Bizon92,Glendenning88}.}.
Their stability is described below.

\begin{figure}[t]%
\includegraphics[width=0.50\textwidth]{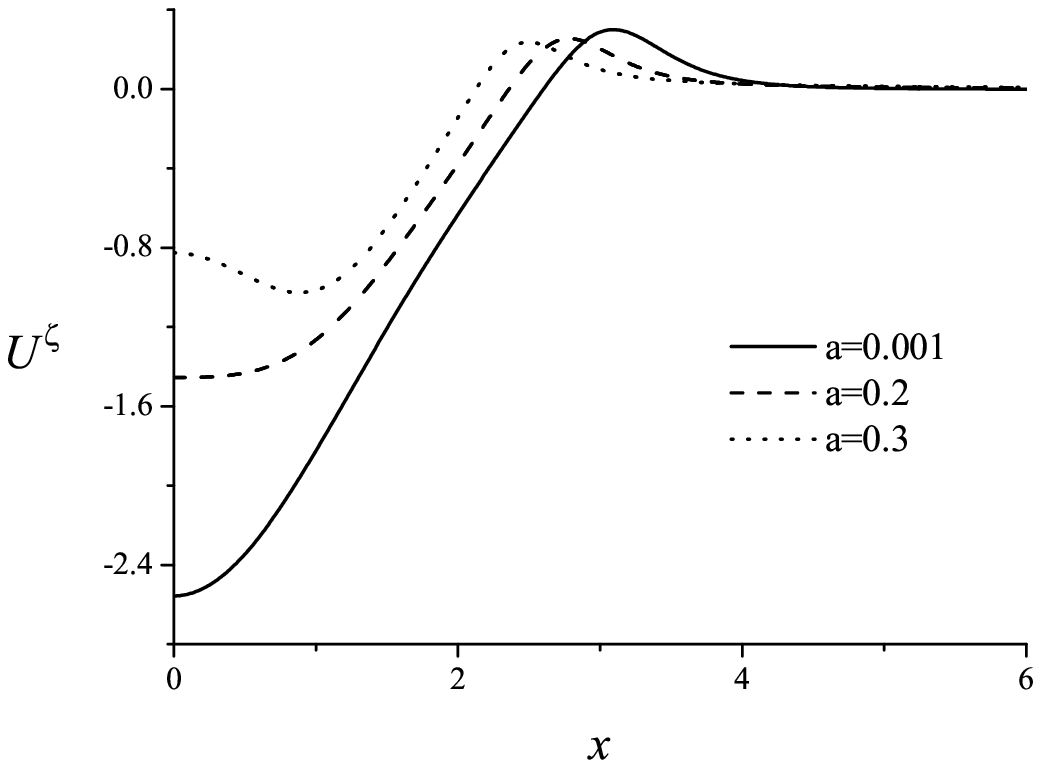}
\includegraphics[width=0.50\textwidth]{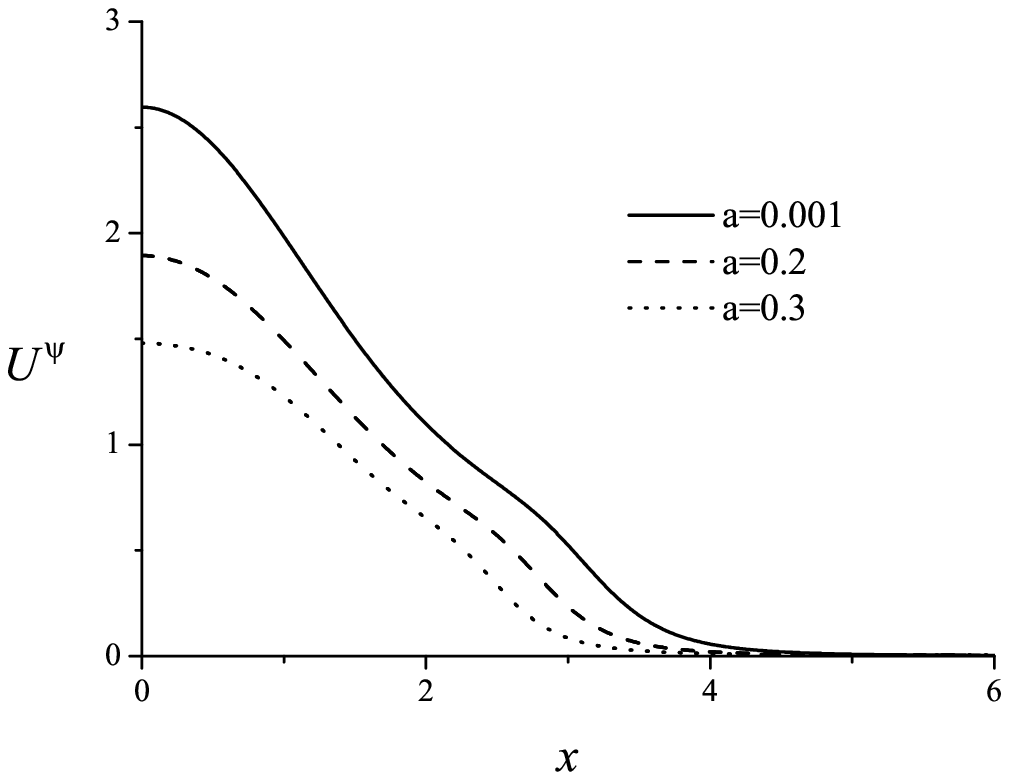}
\caption{The potentials $U^{\zeta}$ and $U^{\Psi}$ of the lower-branch solitons for $D_{\rm eff} =0.00025$ and for several values of $a$.}
\label{fig:potentials_soliton_lower}
\end{figure}

We will start with the lower branch of solutions which is stable for ES solitons, i.e. in the case without scalar field. The so-called potentials
defined by equations (\ref{eq:Potentials_Solitons}) are given in Fig. \ref{fig:potentials_soliton_lower} for some of the soliton
solutions which belong to the lower branch in Fig. \ref{fig:phase_soliton}. As it can be seen the potential $U^{\Psi}$ is
positive and cannot lead to instabilities but $U^{\zeta}$ is negative near the origin which means that unstable modes
could exist. The asymptotic value at infinity of $U^{\zeta}$ is zero, and for the chosen parameters, $U^{\Psi}$ tends to
$U^{\Psi\,\infty}=-0.00346$ which means that the mass of the scalar field is $m=0.0589$ and the period of the
oscillation of the tail is $T=107$ according to eq. (\ref{eq:PeriodTail}).

\begin{figure}[t]%
\includegraphics[width=0.50\textwidth]{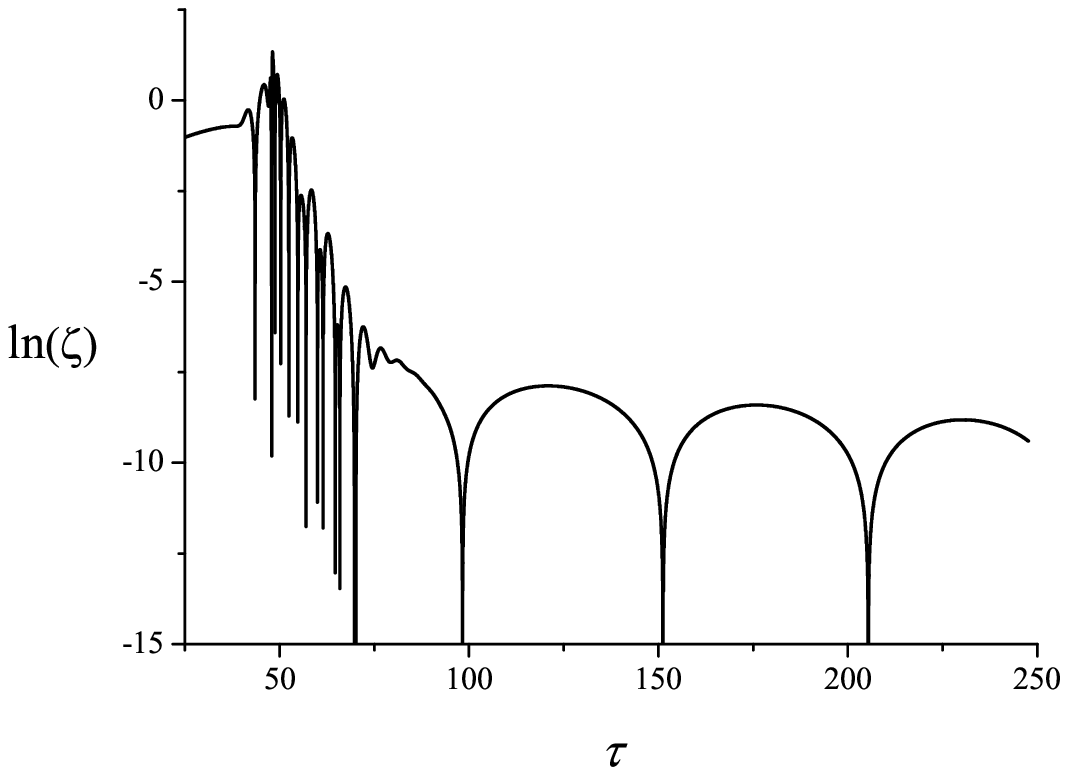}
\includegraphics[width=0.50\textwidth]{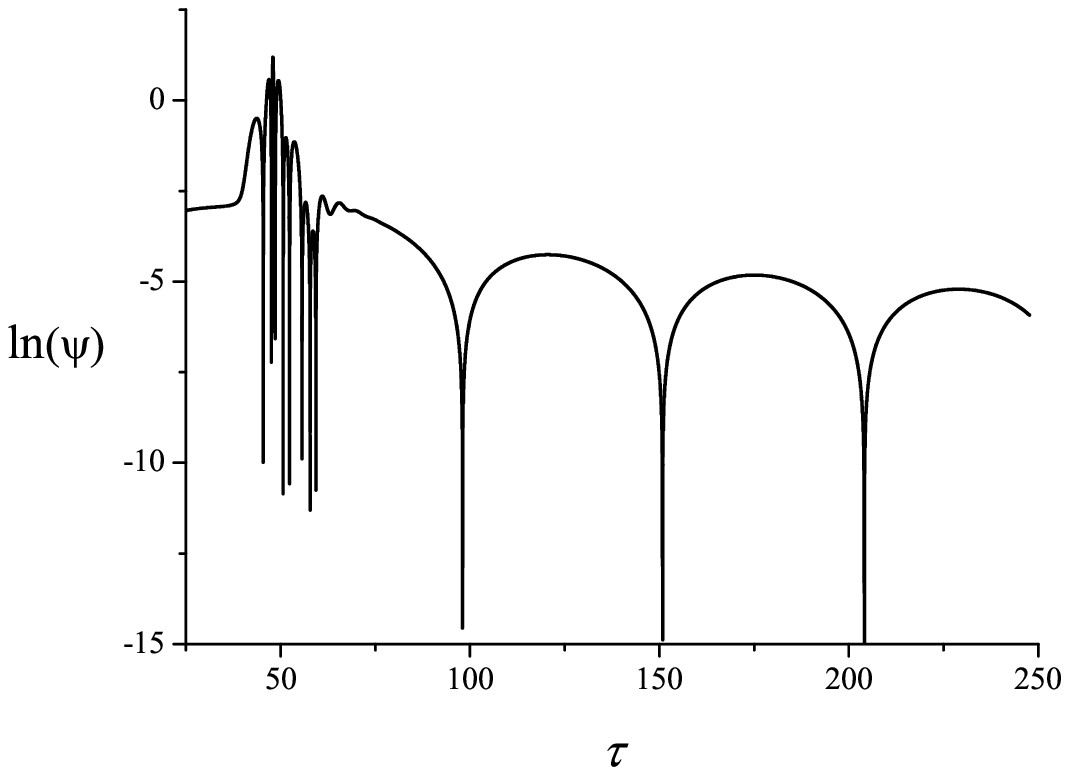}
\caption{The time evolution of the perturbations $\zeta$ and $\Psi$ of a lower-branch soliton solution with $D_{\rm eff} =0.00025$ and $a=0.2$.}
\label{fig:wave_functions_soliton_lower}
\end{figure}

We evolve the coupled wave equations (\ref{eq:wave_like_zeta})--(\ref{eq:wave_like_psi}) with the appropriate
QNM boundary conditions -- the perturbations should be regular at the origin $x=0$ (i.e. in our case $\zeta_{x=0}=0$ and $\Psi_{x=0}=0$) and have the form of an outgoing
wave at infinity. It turns out that all of the studied solutions which belong to the lower branch 
are stable against  the considered perturbations. The time evolution of a Gaussian initial perturbation is presented in
Fig. \ref{fig:wave_functions_soliton_lower}. The wave form consists of quasinormal oscillations in 
early times and an oscillatory tail for late times, where the period of the tail oscillations is the same as the period
predicted by eq. (\ref{eq:PeriodTail}) within numerical errors.

\begin{figure}[t]%
\includegraphics[width=0.50\textwidth]{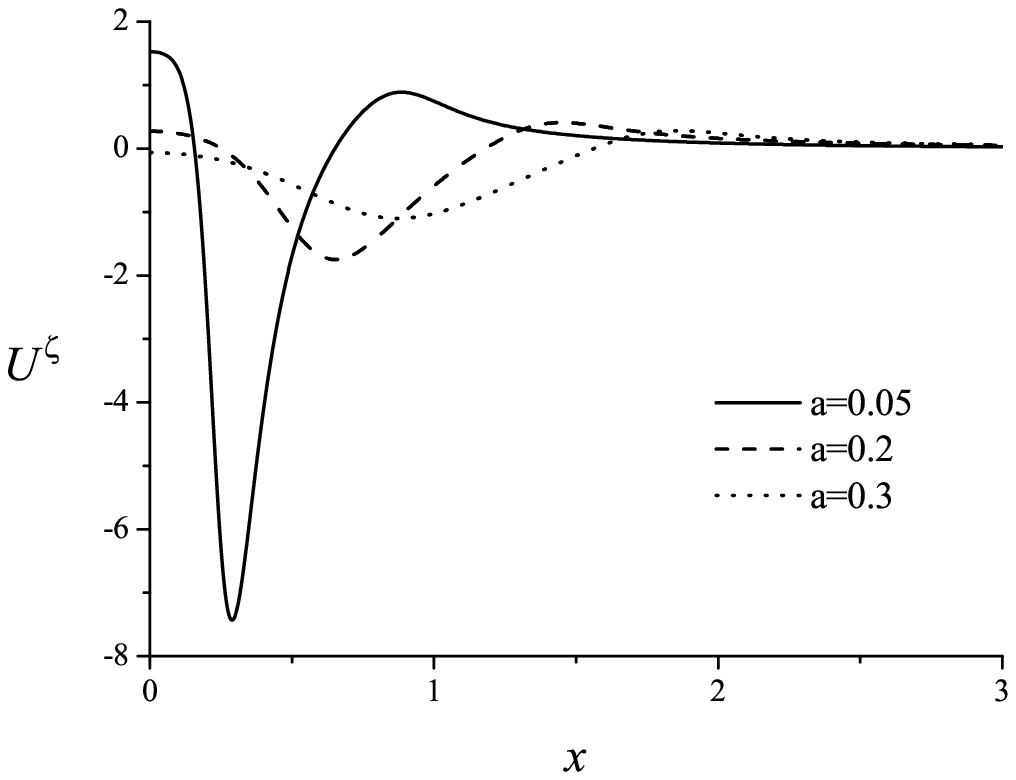}
\includegraphics[width=0.50\textwidth]{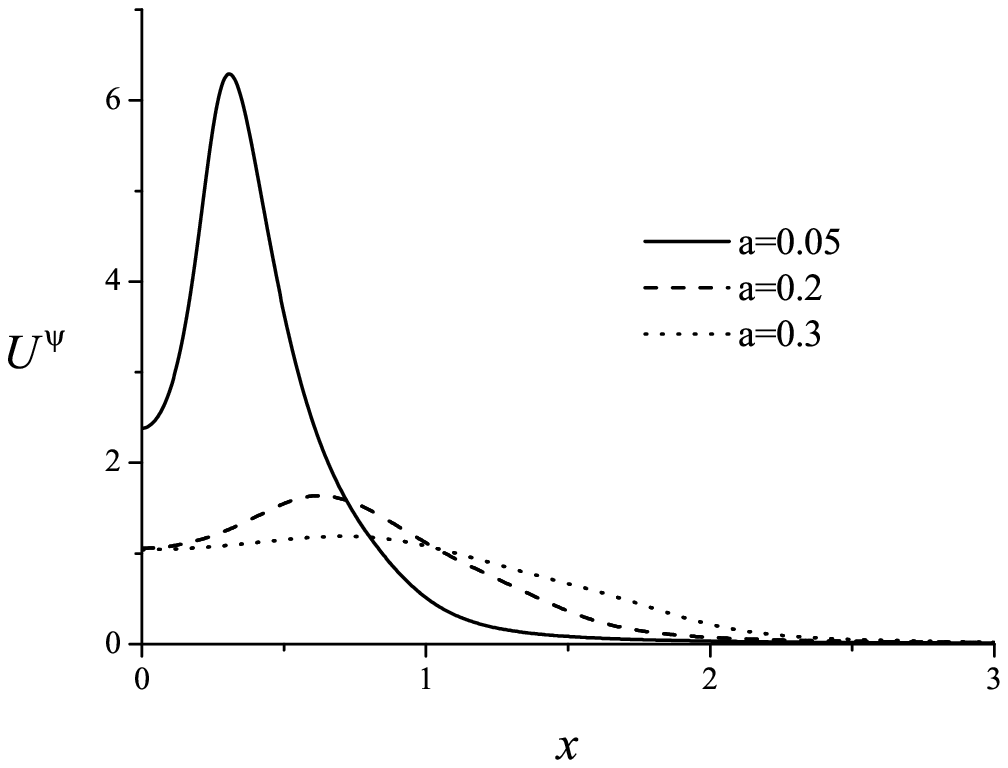}
\caption{The potentials $U^{\zeta}$ and $U^{\Psi}$ of the upper-branch solitons for $D_{\rm eff} =0.00025$ and for several values of $a$.}
\label{fig:potentials_soliton_upper}
\end{figure}

\begin{figure}[t]%
\includegraphics[width=0.50\textwidth]{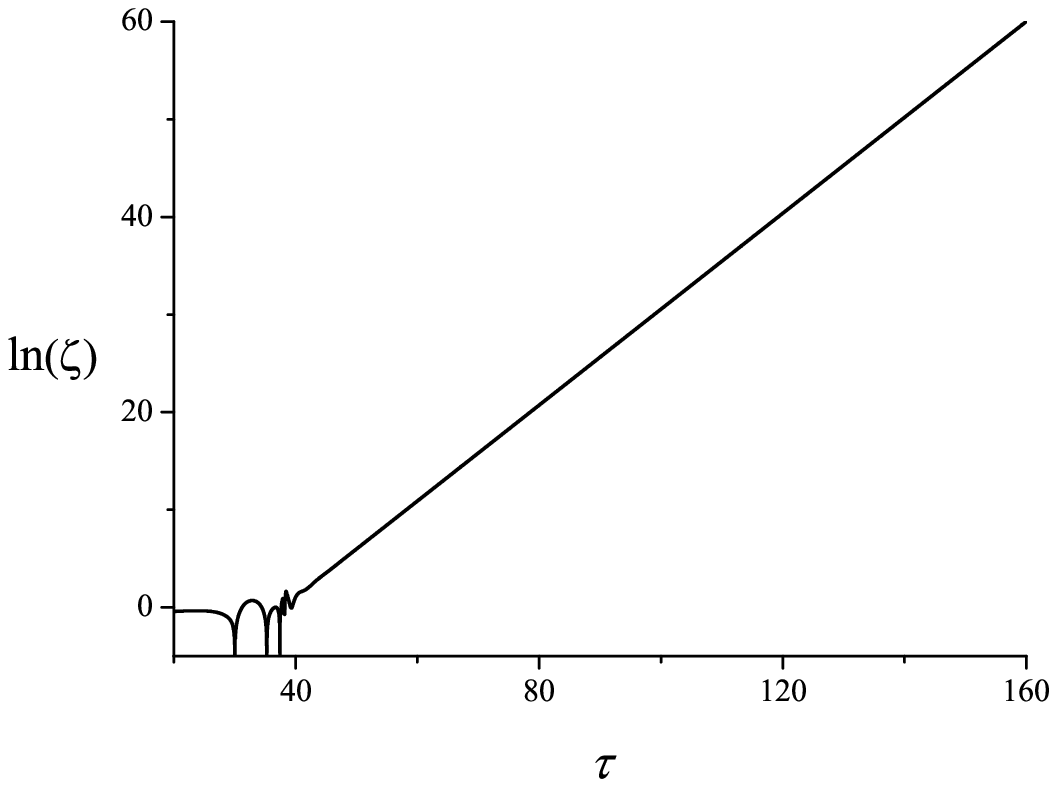}
\includegraphics[width=0.50\textwidth]{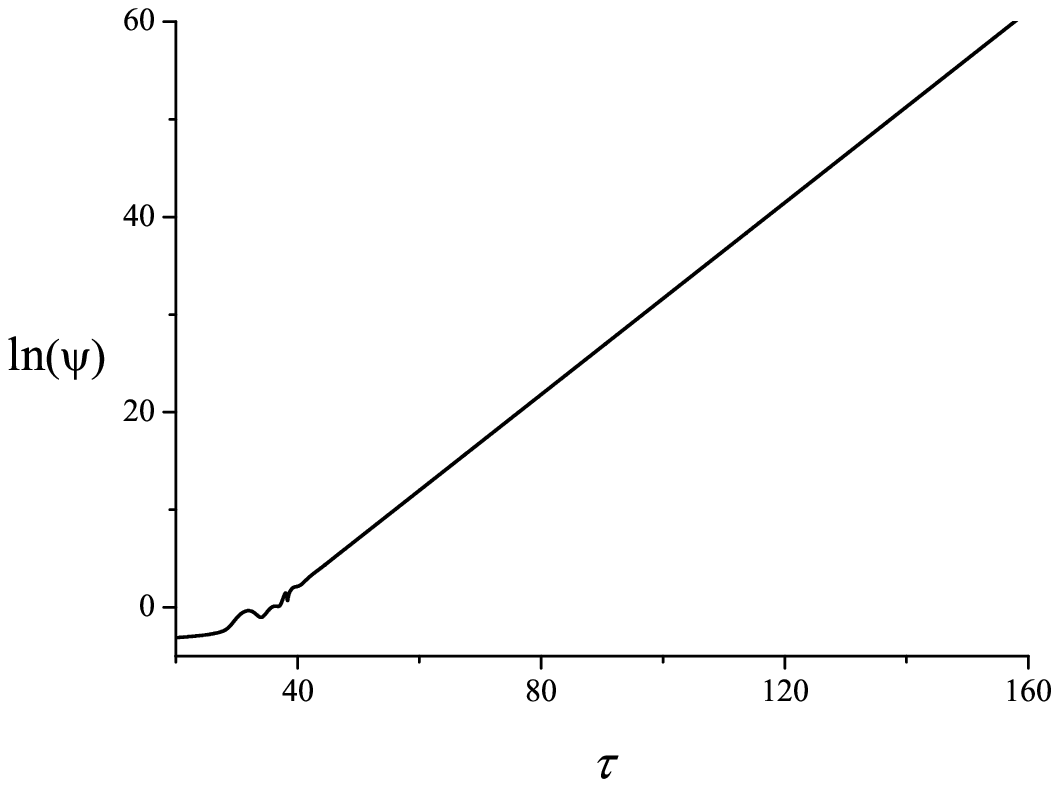}
\caption{The time evolution of the perturbations $\zeta$ and $\Psi$ of an upper-branch soliton solution with $D_{\rm eff} =0.00025$ and $a=0.2$.}
\label{fig:wave_functions_soliton_upper}
\end{figure}

Now let us consider the upper branch of soliton solutions which is unstable for the ES solitons.
The two potentials $U^{\zeta}$ and $U^{\Psi}$ are shown in Fig. \ref{fig:potentials_soliton_upper}. Again $U^{\Psi}$
is positive and $U^{\zeta}$ is negative near the origin but on this branch the negative part  is deeper than on the lower branch.
This may lead to instabilities in the wave equation for $\zeta$, which will also affect the perturbations of the scalar field $\Psi$ through the
coupling terms. When we evolve equations (\ref{eq:wave_like_zeta})--(\ref{eq:wave_like_psi}), it turns out that all
of the studied upper-branch solutions are unstable. The time evolution of a Gaussian initial perturbation is presented in
Fig. \ref{fig:wave_functions_soliton_upper}. As it can be seen, the perturbations $\zeta$ and $\Psi$ grow exponentially
with time, i.e. the QNM frequencies are purely imaginary.

\begin{figure}[t]%
\includegraphics[width=0.50\textwidth]{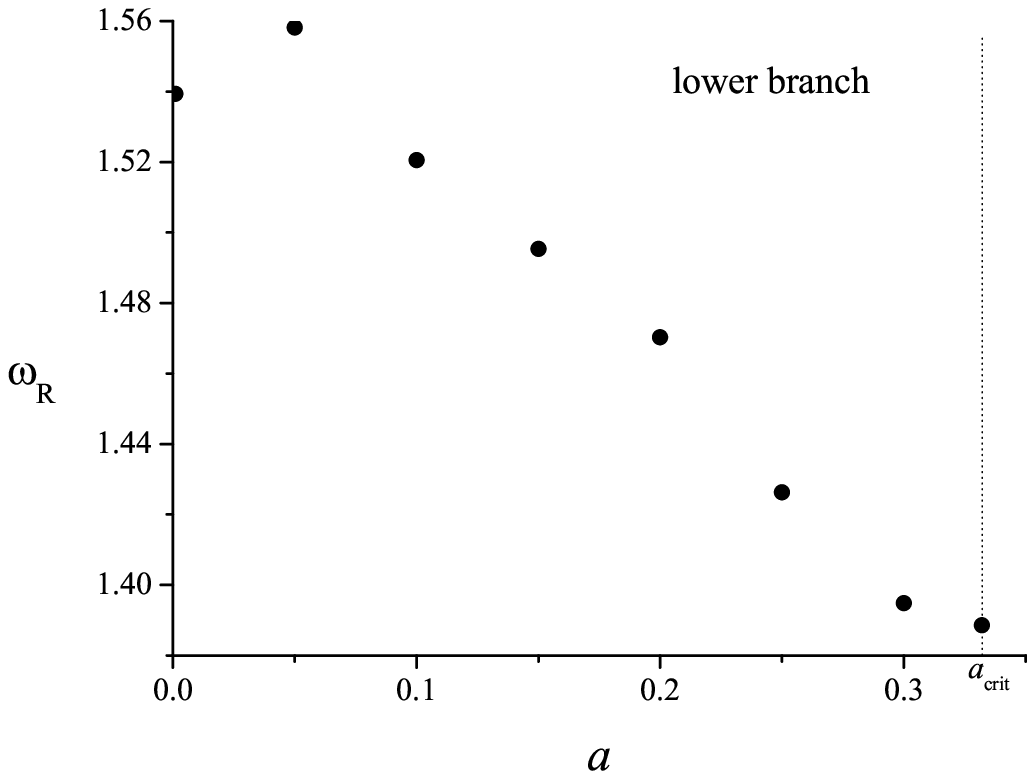}
\includegraphics[width=0.50\textwidth]{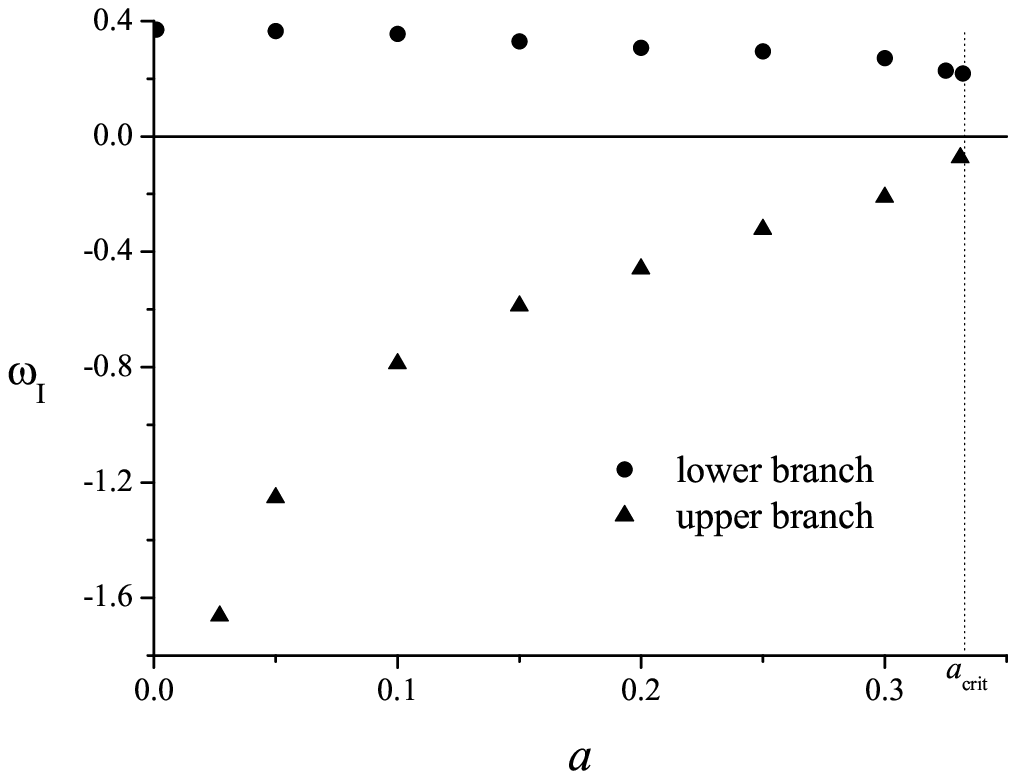}
\caption{The real (left panel) and the imaginary (right panel) parts of the frequencies as a function of the parameter $a$ for the lower and the upper branches of soliton solutions  ($D_{\rm eff} =0.00025$).
The frequencies are obtained using the time evolution.}\label{fig:soliton_freqs}
\end{figure}

The calculated frequencies for both of the branches are shown in Fig. \ref{fig:soliton_freqs}. The error of the
obtained frequencies is big and can reach up to $20\%$ in some cases. The reason for this is the more complicated  wave form which is
due to the coupling of the wave equations and the presence of mass term in the wave equation for the perturbations of the scalar field $\Psi$. Also, the
background solutions are known only numerically, which is an additional complication. But even if we take into account the
error, the numerical values of the QNM frequencies differ significantly from the case without a scalar field \cite{Zajac09} (the absolute
values of the real and the imaginary parts of the QNM frequencies can be several times bigger here than in the ES case).

The qualitative behavior of the frequencies as we vary the parameter $a$, which is shown in Fig. \ref{fig:soliton_freqs}, is
the one we expected from the case without a scalar field \cite{Zajac09}.
When we increase the value of the parameter $a$, the real $\omega_R$ and the imaginary $\omega_I$ parts of the frequencies of the stable lower
branch decrease,
while the frequencies $\omega_I$ of the unstable modes of the upper branch
increase (as we already said $\omega_R=0$ for the unstable upper branch). In the limit $a\rightarrow a_{\rm crit}$
(i.e. when we approach the value of the parameter $a$ where the two branches merge), the $\omega_I$ of the upper and the
lower branches tend to zero, i.e. they indicate a stability change.

When we increase the value of the parameter $D_{\rm eff}$ the errors of the obtained QNM frequencies increase
mainly because the background solutions become more difficult to obtain and the change in the
frequencies as we vary $D_{\rm eff}$ is within numerical errors.

\subsection{Black Holes}\label{sec:Black_Holes}
\begin{figure}
    \begin{center}
    \includegraphics[width=7cm]{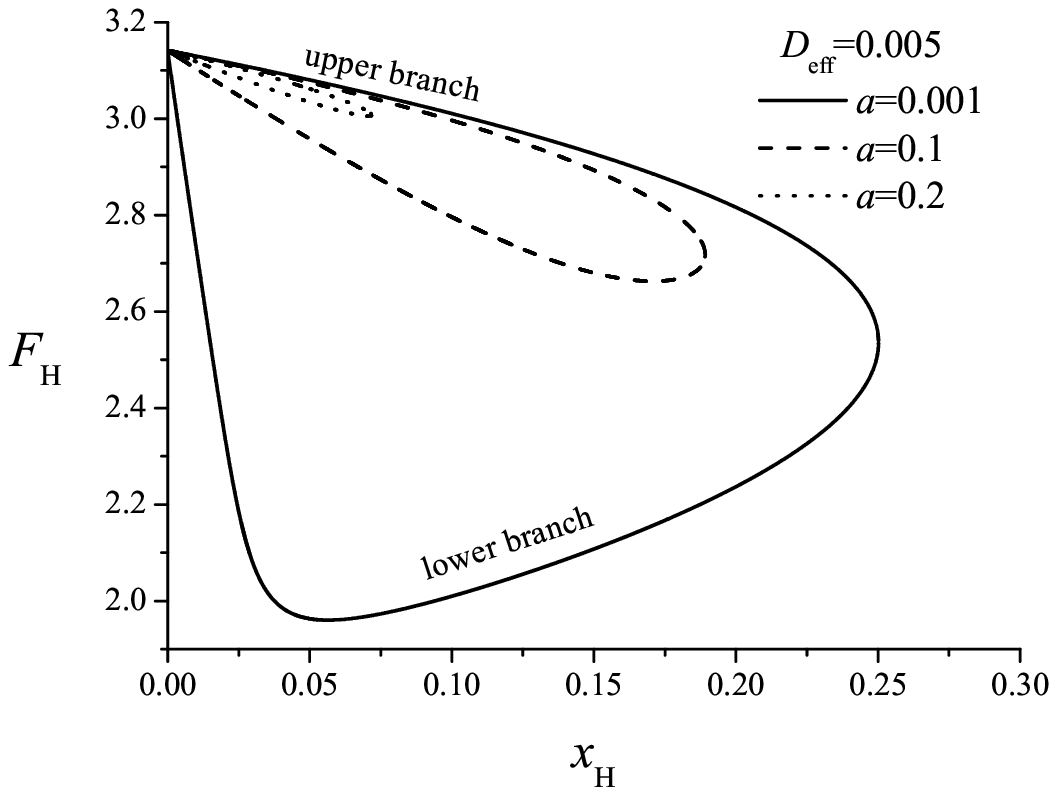}
    \includegraphics[width=7cm]{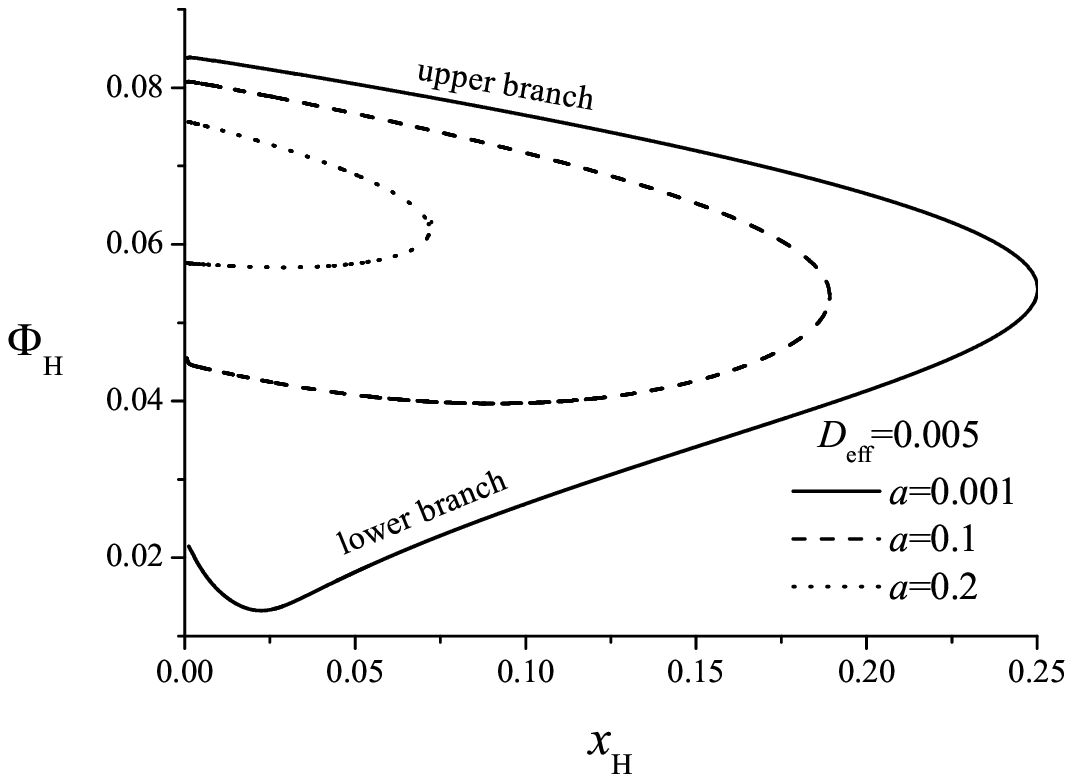}
    \caption{The $F_{H}(x_H)$ and $\Phi_{H}(x_H)$ phase diagrams for sequences of black-hole solutions for $D_{\rm eff}=0.005$ and for
    different values of $a$. }
    \label{fig:phase_black_hole}
\end{center}
\end{figure}

Because of the presence of an event horizon, the Skyrme black holes are topologically trivial. The shooting parameters for the background
black-hole solutions are the values of the Skyrmion and the scalar fields at
the horizon $x_H$ -- $F_H$ and $\Phi_H$, respectively. The $F_H(x_H)$ and $\Phi_H(x_H)$ phase diagrams for sequences of black-hole
solutions, obtained in \cite{DSY_SKYRME}, are shown in Fig. \ref{fig:phase_black_hole}. As we can see, again two
branches of solutions exist (upper and lower) which merge at some critical value of the radius of the horizon $r_{\rm H \,
crit}$ \footnote{The upper and the lower branches are defined using the $F_H(x_H)$ diagram. This is obviously different from the soliton
case, and actually, the upper branch for black holes will have properties (such as stability and finiteness/divergency of
some of the functions as $a \rightarrow 0$) similar to the lower branch for solitons and vice versa.}. The stability of the two branches is described below.

\begin{figure}[t]%
\includegraphics[width=0.50\textwidth]{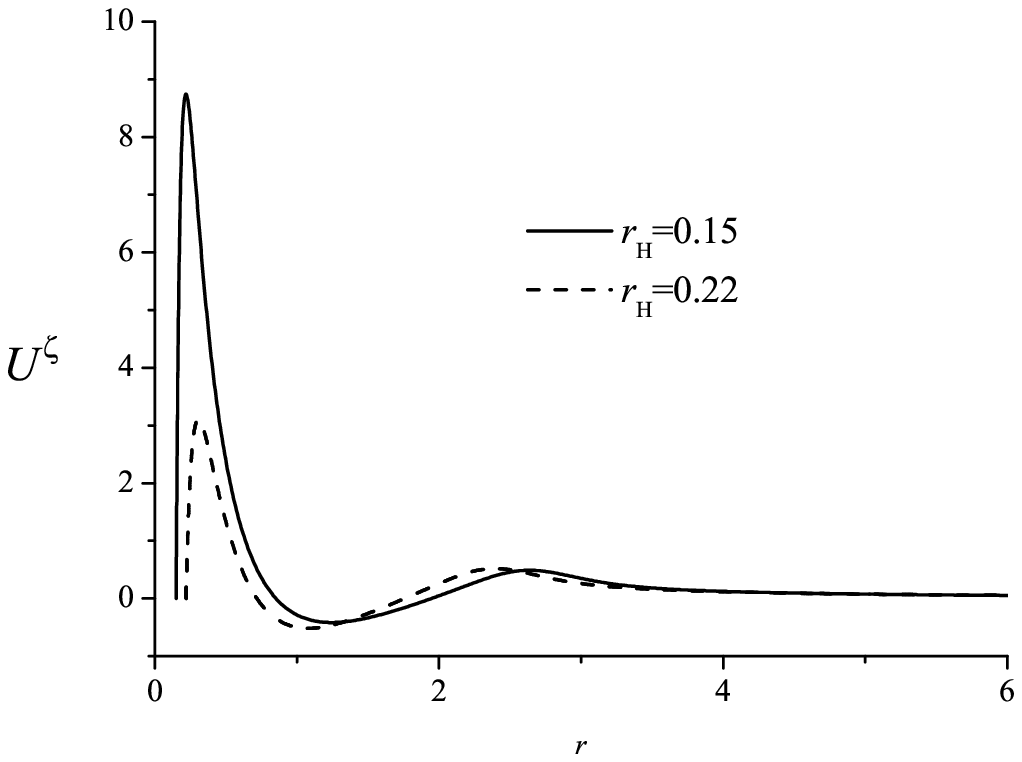}
\includegraphics[width=0.50\textwidth]{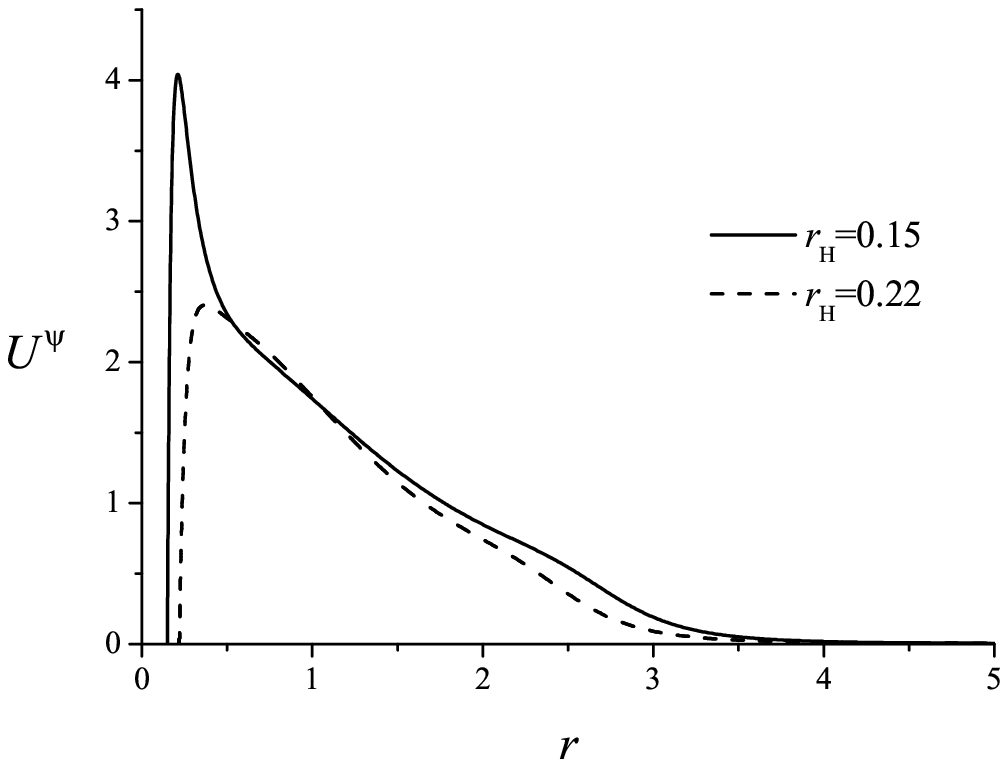}
\caption{The potentials $U^{\zeta}$ and $U^{\Psi}$ of the upper-branch black holes for $D_{\rm eff} =0.00025$ and for $a=0.1$.}
\label{fig:potentials_black_hole_upper}
\end{figure}

\begin{figure}[t]%
\includegraphics[width=0.50\textwidth]{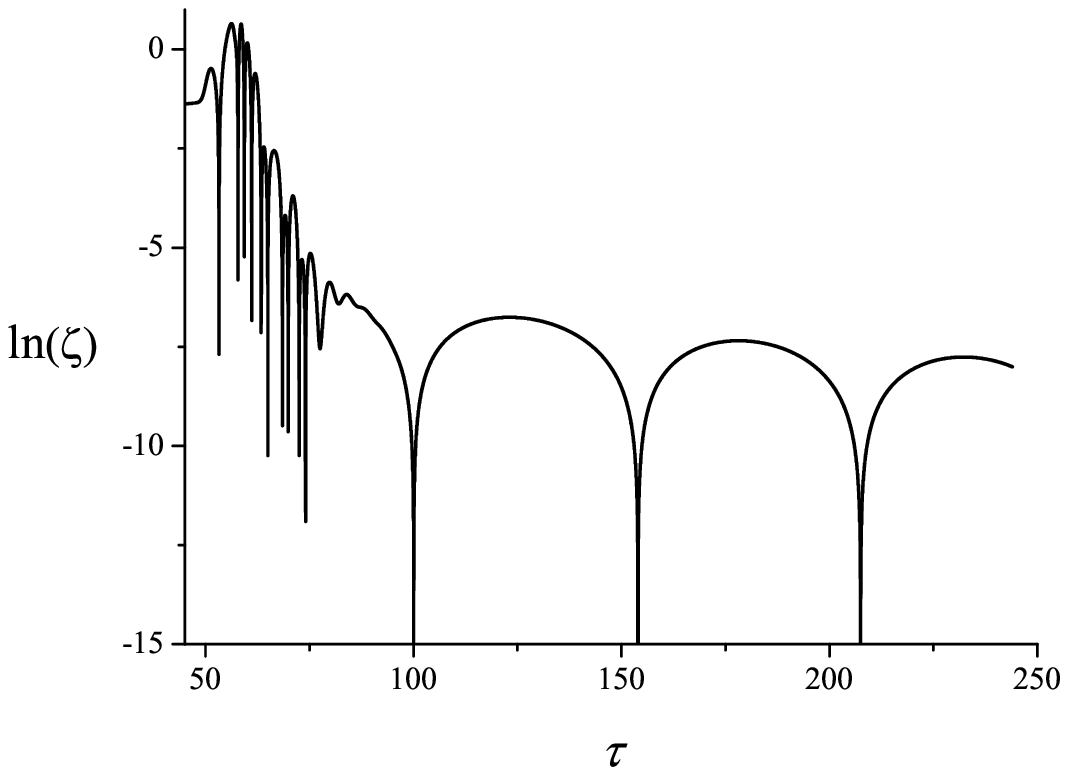}
\includegraphics[width=0.50\textwidth]{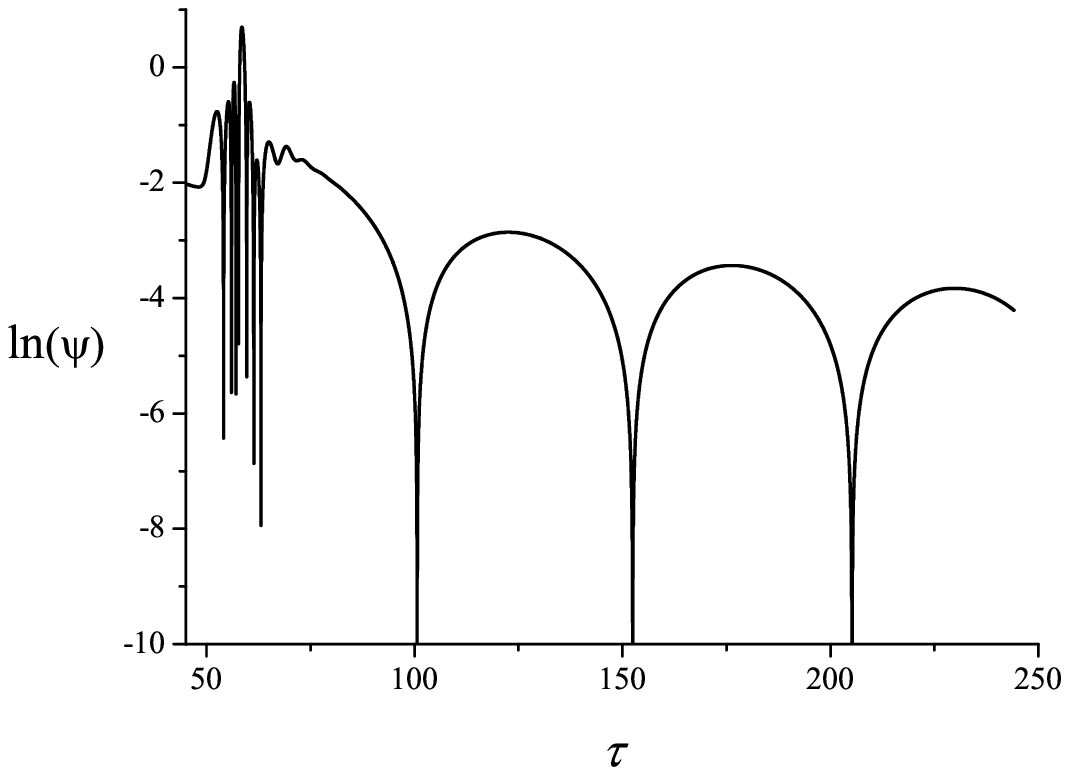}
\caption{The time evolution of the perturbations $\zeta$ and $\Psi$ of an upper-branch black-holes solution with
$D_{\rm eff} =0.00025$, $a=0.1$, and $r_H=0.1$.}
\label{fig:wave_functions_black_hole_upper}
\end{figure}

We will start with the upper branch which is stable for the ES black holes, i.e. in the case without a scalar field. The potentials
$U^{\zeta}$ and $U^{\Psi}$ for some of the upper-branch solutions are given in Fig. \ref{fig:phase_black_hole}, are given in
Fig. \ref{fig:potentials_black_hole_upper}. Similar to the soliton case, the potential $U^\Psi$ is positive and
$U^\zeta$ has a negative minimum near the horizon. We evolve equation (\ref{eq:wave_zeta})--(\ref{eq:wave_psi})
with the standard boundary conditions -- purely ingoing waves at the horizon and purely outgoing waves at
infinity. It turns out that all of the studied black holes of the upper branch are stable against the considered perturbations and the
time evolution of a Gaussian perturbation is shown in Fig. \ref{fig:wave_functions_black_hole_upper}. Again, two stages
of the time evolution are observed -- the quasinormal ringing and the oscillatory tail with a period given by eq.
(\ref{eq:PeriodTail}).

\begin{figure}[t]%
\includegraphics[width=0.50\textwidth]{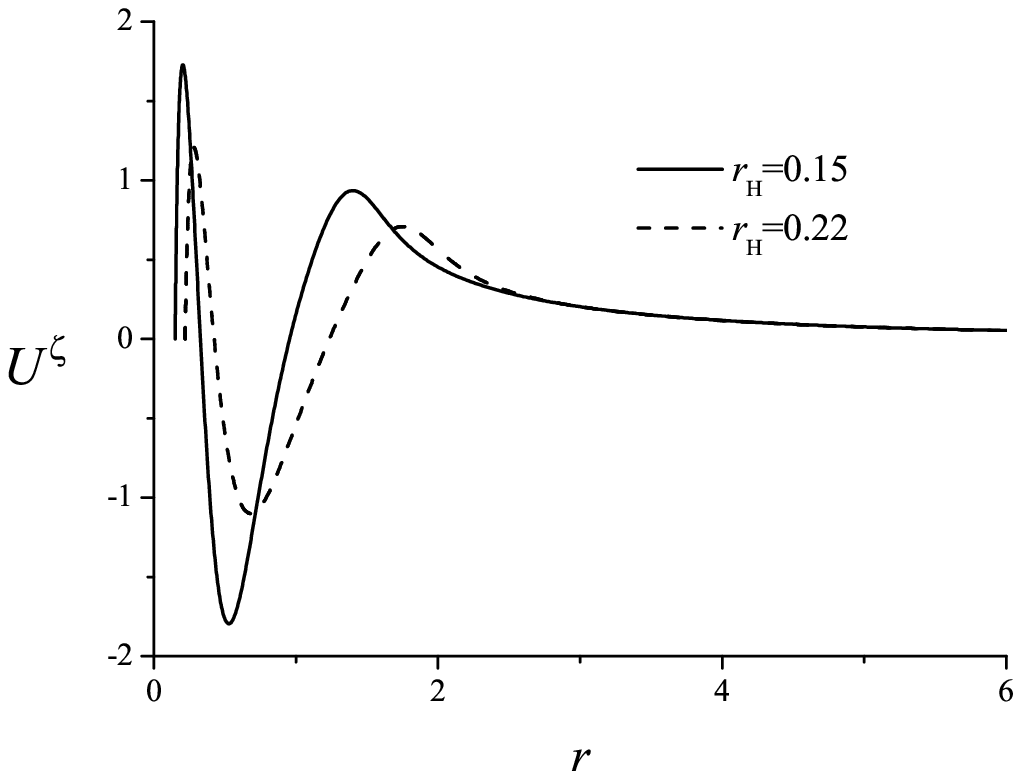}
\includegraphics[width=0.50\textwidth]{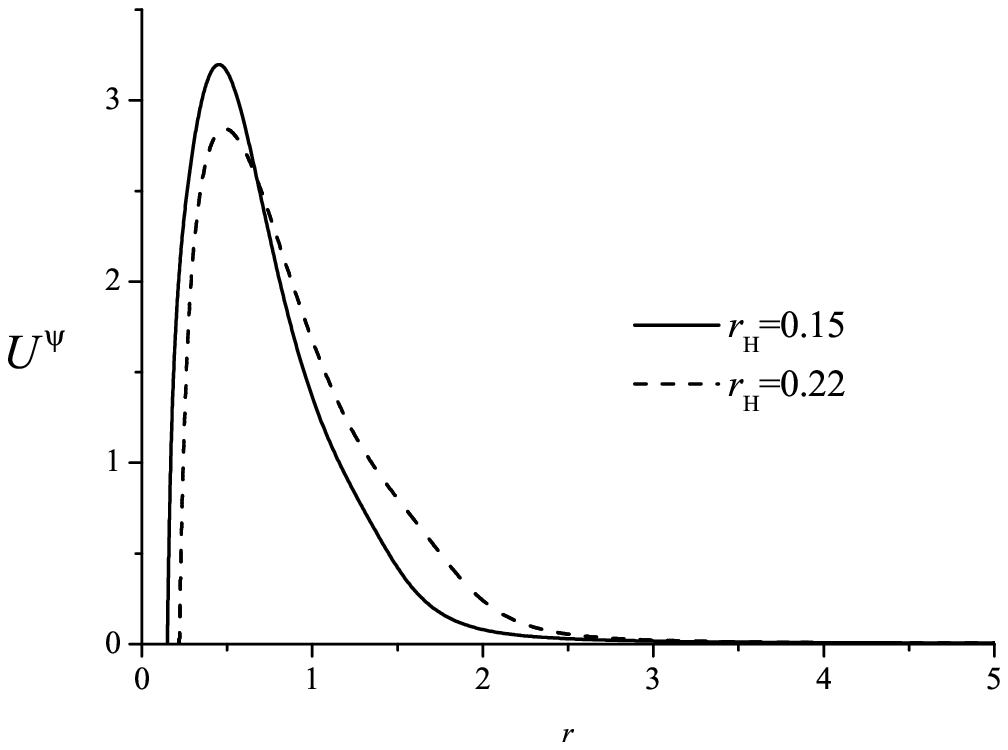}
\caption{The potentials $U^{\zeta}$ and $U^{\Psi}$ of the lower-branch black holes for $D_{\rm eff} =0.00025$ and for $a=0.1$.}
\label{fig:potentials_black_hole_lower}
\end{figure}

\begin{figure}[t]%
\includegraphics[width=0.50\textwidth]{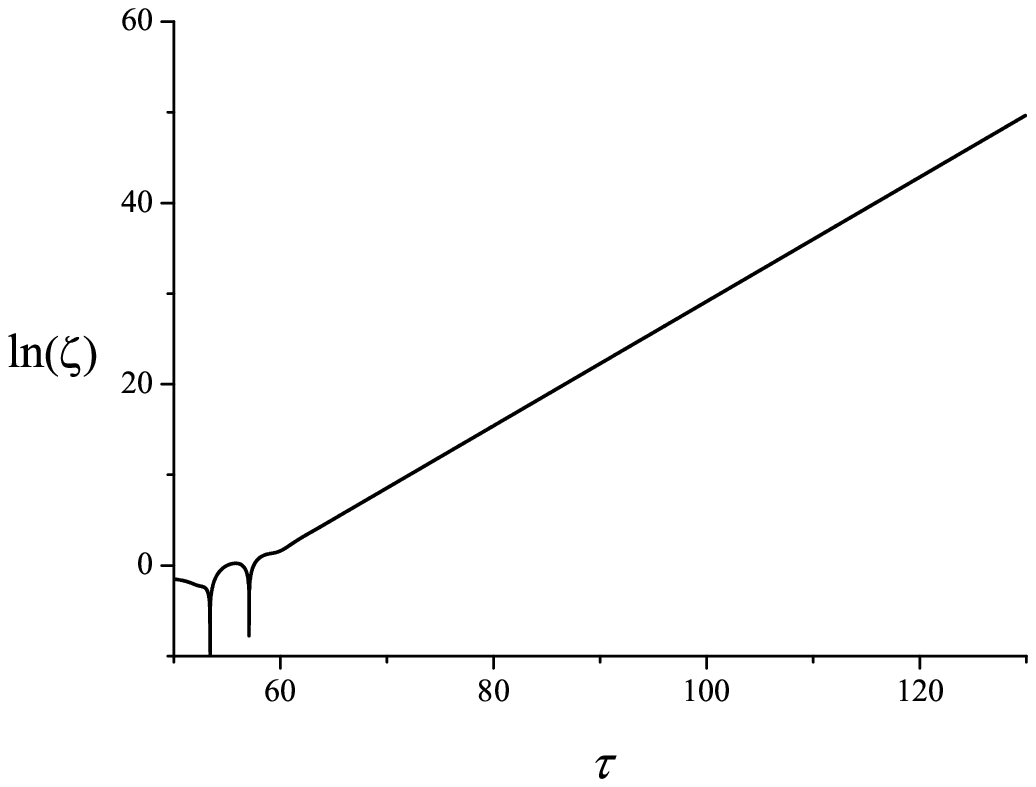}
\includegraphics[width=0.50\textwidth]{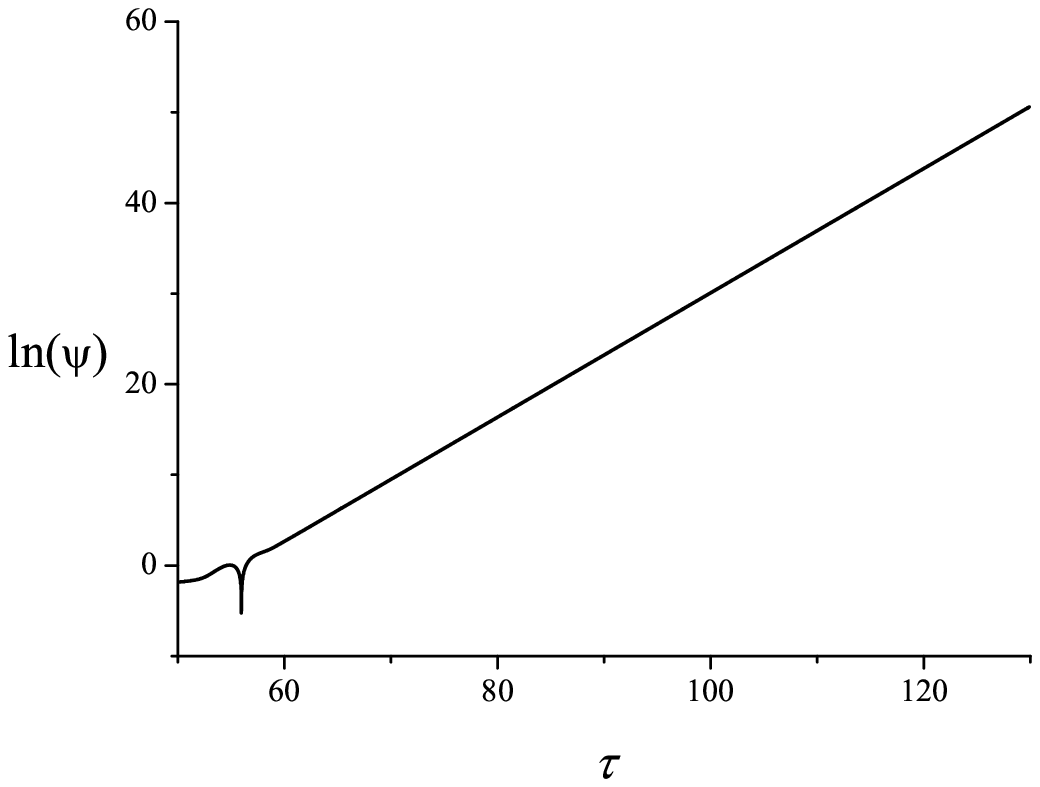}
\caption{The time evolution of the perturbations $\zeta$ and $\Psi$ of a lower-branch black-hole solution with $D_{\rm eff} =0.00025$, $a=0.1$, and $r_H=0.1$.}
\label{fig:wave_functions_black_hole_lower}
\end{figure}

The potentials $U^{\zeta}$ and $U^{\Psi}$ for some of the solutions which belong to the lower branch are shown in Fig.
\ref{fig:potentials_black_hole_lower} \footnote{The lower branch is unstable for the ES black holes, i.e. in the case without a scalar field}.
$U^{\Psi}$ is again positive but $U^{\zeta}$ has a negative minimum near the
horizon which is much deeper here than for the corresponding black holes of the upper branch, which could lead to
instabilities. Indeed the time evolution shows  that all of the solutions of the lower branch are
unstable. The logarithm of the wave functions are shown in Fig. \ref{fig:wave_functions_black_hole_lower}, where both $\zeta$ and $\Psi$ grow
exponentially with time.

\begin{figure}[t]%
\includegraphics[width=0.50\textwidth]{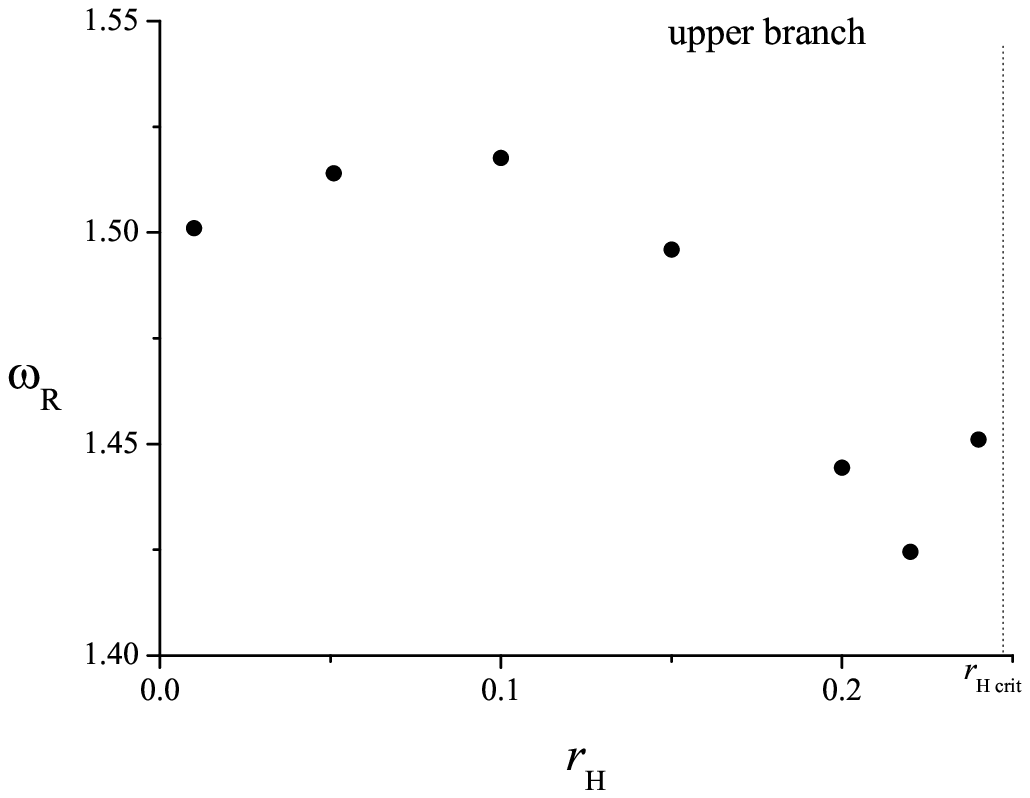}
\includegraphics[width=0.50\textwidth]{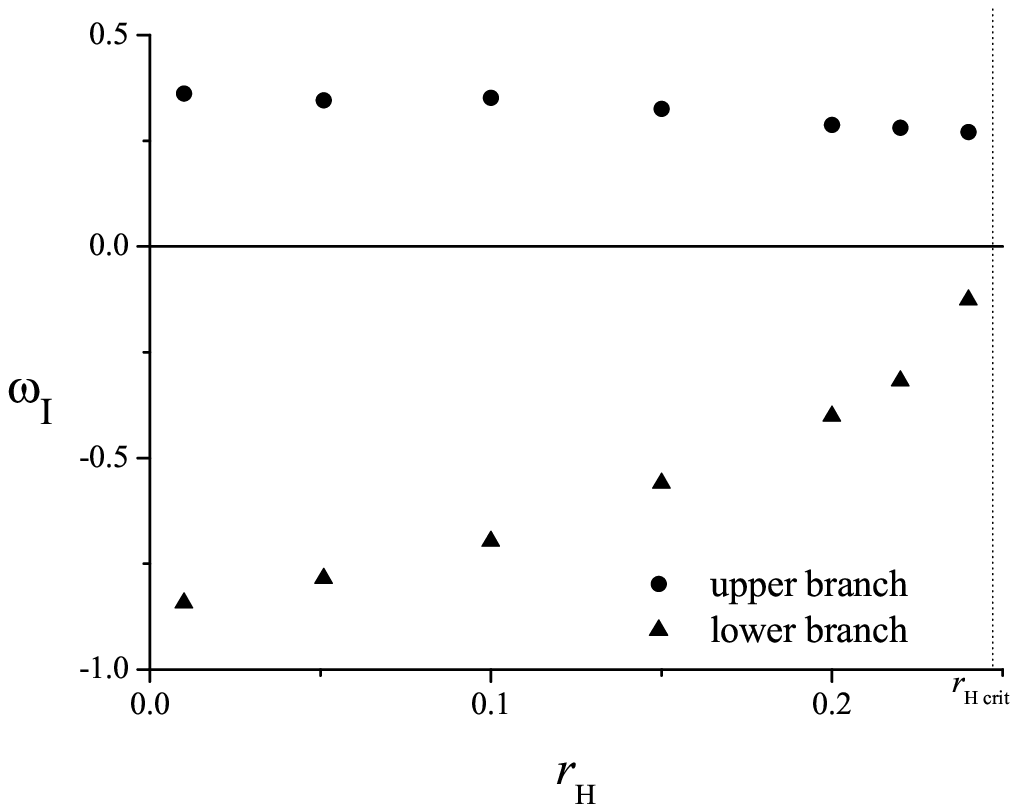}
\caption{The real (left panel) and the imaginary (right panel) parts of the frequencies as a function of the radius of the horizon $x_H$ for the
lower and the upper branches of black-hole solutions
 ($D_{\rm eff} =0.00025$ and $a=0.1$). The frequencies are obtained using the time evolution.}\label{fig:black_hole_freqs}
\end{figure}

The corresponding QNM frequencies for the upper and the lower branches are shown in Fig. \ref{fig:black_hole_freqs}.
The error is again big (can reach up to $20\%$) but we can comment on the qualitative behavior of the frequencies.
The real and the imaginary parts of the stable upper-branch frequencies decrease when  we approach the critical
value $x_{\rm H \, crit}$ where the two branches merge. On the lower branch, which is unstable, the  imaginary part of the
frequencies increases when we increase $x_H$ ($\omega_R=0$ for this branch). So, on both branches, in the limit
$x_H \rightarrow x_{\rm H \, crit}$, the imaginary parts of the frequencies $\omega_I$  become zero, i.e. a change of stability is
observed.

As discussed in \cite{DSY_SKYRME}, the properties of the black-hole solutions when we vary $a$ for fixed $x_H$ are
similar to those of the solitons -- two branches of black holes exit which merge at some critical value of the parameter $a_{\rm
crit}$. The  behavior of the QNM frequencies is also similar to the one shown in Fig.
\ref{fig:soliton_freqs} for the soliton case. With the increase of the parameter $a$ the real and the imaginary parts of the frequencies of
the stable branch decreases, and the imaginary part of the frequencies of the unstable branch increases. The imaginary
parts of both the stable and the unstable branches of black holes tend zero when $a_{\rm crit}$ is approached.
Similar to the soliton case, the change in the frequencies when $D_{\rm eff}$ varies is within  numerical errors.

\section{Summary of the results}\label{sec:Conclusion}
As the results indicate, the dilaton does not change the stability of the solutions. Again, the so-called lower branch
of solitons is stable against radial perturbations while the upper is unstable. For the black holes the  upper branch
is stable and the lower is unstable.  The thermodynamical stability analysis of the black holes presented in
\cite{DSY_SKYRME}, which is based on the turning point method, is also in agreement with the results from the linear stability analysis.

The modes of the unstable solutions are purely imaginary, i.e   $\omega_R=0$ and $\omega_I<0$. The modes of the stable solutions
are damped oscillations with $\omega_R\neq0$ and $\omega_I>0$. Hence, at the point of stability change  both the imaginary part and the real
part of the QNM frequencies become zero. The time evolution of the black-hole solutions in the vicinity
of the point where the two branches merge could not be studied accurately, but still the results presented in
Figs. \ref{fig:soliton_freqs} and \ref{fig:black_hole_freqs}, for the solitons and the black holes, respectively, show the expected
qualitative behavior. Also, the numerical values of the QNM frequencies can be significantly different from the case without a
scalar field, i.e. the presence of a scalar field significantly alters  the spectrum quantitatively.

One more interesting observation can be made. If we consider the evolution of a test scalar field on both black-hole
and soliton backgrounds, it turns out that all of the modes are damped, i.e. they are stable for both branches -- the stable and the unstable ones.
Thus the time evolution of the test scalar field cannot be used to study the stability of the branches.
\vskip 1cm

\section*{Acknowledgements}
This work was partially supported by the Bulgarian National Science Fund under Grants DO 02-257, by Sofia
University Research Fund under Grant No 88/2011. D.D. would like to thank the DAAD for the support, the Institute
for Astronomy and Astrophysics T\"{u}bingen for its kind hospitality. D.D. is also supported by the German Science Council (DFG) via SFB/TR7. D.D. would like to thank Paul Lasky for valuable discussions.

\appendix
\section{Coefficients in the wave equations}\label{sec:coefficients}
The coefficients in the wave equations Eqs. (\ref{eq:wave_like_zeta}) and (\ref{eq:wave_like_psi}) are
\begin{eqnarray}
A_1&=&{\chi_0\,'-\alpha_0\,'\over2}\frac {u_{0F} F_0^{\,'} }{u_0 }
-{a\over x}{e^{\alpha_0 }}v_{0F}F_0^{\,'}
-{a^2\over 2 x^2}e^{\alpha_0 }u_0 v_0 F_0^{\,'2}\notag\\
&-&{1\over 2 }{\chi_0\,'-\alpha_0\,'\over2}\frac {u_{0}' }{u_0 }
+{ a\over 2 } e^{\alpha_0 }u_0 F_0^{\,'2}\tilde{V}_0
+\frac {u_{0F}' F_0^{\,'} }{u_0 }+\frac {u_{0F} F_0^{\,''} }{u_0 }\notag\\
&+&{1\over 4 }\frac {u_{0}^{\,'2} }{u_0^2 }
-{1\over 2 }\frac {F_0^{\,'2}u_{0FF}}{u_0 }
-{1\over 2 }e^{\alpha_0 }\frac {v_{0FF}}{u_0 }
-{1\over 2 }\frac {u_{0}^{\,''} }{u_0 }
+{a\over x^2}e^{\alpha_0 }u_0 F_0^{\,'2},
\end{eqnarray}
\begin{equation}
A_2={u_{0\Phi} F_0^{\,'}\over x \sqrt{u_0}},
\end{equation}
\begin{eqnarray}
A_3&=&-{1\over 2}e^{\alpha_0}{\frac {v_{0F\Phi} }{x\sqrt {u_0 }}}
+{\chi_0\,'-\alpha_0\,'\over2}{\frac {  u_{0\Phi} }{x\sqrt {u_0 }}}F_0^{\,'}
-{\frac {u_{0\Phi} }{{x}^{2}\sqrt {u_0 }}}F_0^{\,'}\notag\\
&-&{1\over 2}\,{\frac {u_{0F\Phi} }{x\sqrt {u_0 }}} F_0^{\,'2}
-{1\over 2}a \tilde{N}e^{\alpha_0}{\frac {v_{0F}}{\sqrt {u_0 }}}\Phi_0^{\,'}
+a\tilde{N}e^{\alpha_0}{\frac {\sqrt {u_0 } F_0^{\,'}\Phi_0^{\,'}}{x}}\notag\\
&-&{a\over 2{x}^{2}}e^{\alpha_0} v_{0\Phi} \sqrt {u_0 } F_0^{\,'}
+{1\over 2}a\tilde{N}x e^{\alpha_0}\sqrt {u_0 }\tilde{V}_0 F_0^{\,'}\Phi_0^{\,'}
+{\frac { u_{0\Phi} }{x\sqrt{u_0 }}} F_0^{\,''}\notag\\
&+&{1\over 2}e^{\alpha_0}\sqrt {u_0} \tilde{V}_{0\Phi} F_0^{\,'}
-{{a}^{2}\tilde{N}\over 2 x}e^{\alpha_0}\sqrt {u_0 } v_0 F_0^{\,'}\Phi_0^{\,'} + \frac{u_{\Phi}^{\,'}}{\sqrt{u_0}x} F_0^{\,'}
\end{eqnarray}
\begin{eqnarray}
B_1&=&-{1 \over 2}\,{\frac {{{\rm e}^{\alpha}}v_{0\Phi\Phi}}{\tilde{N}{x}^{2}}}-{{\chi_0\,'-\alpha_0\,'\over 2x}}
-{1 \over 2}\,{\frac { u_{0\Phi\Phi}}{\tilde{N}{x}^{2}}}F_0^{\,'2} \notag\\
&+& x{{\rm e}^{\alpha}} \tilde{V}_{0 \Phi} \Phi_0^{\,'} +
{1 \over 2}\,{\frac {{{\rm e}^{\alpha }}\tilde{V}_{0 \Phi\Phi}}{a\tilde{N}}}+
{a\tilde{N}{\rm e}^{\alpha}}\Phi_0^{\,'2}-{\frac {a{\rm e}^{\alpha} v_{0\Phi}}{x}}\Phi_0^{\,'}\notag\\
&-&{a^2 \tilde{N} \over 2}\,{{\rm e}^{\alpha}}v_0\Phi_0^{\,'2} +
{a \tilde{N} \over 2}\,{x}^{2} {{\rm e}^{\alpha}} \tilde{V}_{0} \Phi_0^{\,'2}
\end{eqnarray}
\begin{equation}
B_2=-\frac {u_{0\Phi}}{\tilde{N} x\sqrt{u_0}}F_0^{\,'}
\end{equation}
\begin{eqnarray}
B_3&=&{1 \over 2}\,\frac{\sqrt {u_0}{\rm e}^{\alpha}\tilde{V}_{0 \Phi}}{\tilde{N}}F_0^{\,'} -
{a^2 \over 2}\,\frac {{{\rm e}^{\alpha}} v_0 \sqrt{u_0}}{x}F_0^{\,'}\Phi_0^{\,'}+{\frac{a{{\rm e}^{\alpha }}\sqrt {u_0}}{x}} F_0^{\,'}\Phi_0^{\,'} \notag\\
&-&{1 \over 2 \tilde{N}}\,{\frac {u_{0 F\Phi}}{x\sqrt{u_0}}}F_0^{\,'2} -
{a \over 2 \tilde{N}}\,{\frac {{{\rm e}^{\alpha}} v_{0\Phi}\sqrt{u_0}}{x^2 }}F_0^{\,'}-{1 \over 2\tilde{N}}\,
\frac {{\rm e}^{\alpha}v_{0 F\Phi}}{x\sqrt {u_0}} \notag\\
&+&{a \over 2}\, x{{\rm e}^{\alpha }} \tilde{V}_{0}\sqrt{u_0}F_0^{\,'}\Phi_0^{\,'} - {a \over 2}\,\frac {{\rm e}^{\alpha}  v_{0 F}}{\sqrt{u_0}}\Phi_0^{\,'} +
{1 \over 2 \tilde{N}}\,{\frac {u_{0\Phi} u_0}{xu_0^{3/2}}}F_0^{\,'}
\end{eqnarray}
and the coefficients in Eqs. (\ref{eq:wave_zeta}) and (\ref{eq:wave_psi}) are
\begin{eqnarray}
\tilde{A}_1=e^{\chi_0-\alpha_0}A_1,~~\tilde{A}_2=e^{(\chi_0-\alpha_0)/2}A_2,~~\tilde{A}_3=e^{\chi_0-\alpha_0}A_3,\\
\tilde{A}_1=e^{\chi_0-\alpha_0}A_1,~~\tilde{B}_2=e^{(\chi_0-\alpha_0)/2}B_2,~~\tilde{B}_3=e^{\chi_0-\alpha_0}B_3.
\end{eqnarray}


\end{document}